\documentclass[aps,reprint,showpacs,superscriptaddress]{revtex4-2}

\usepackage{amssymb}
\usepackage{natbib}
\usepackage{graphicx}
\usepackage{amsmath}
\usepackage[bookmarks = false]{hyperref}
\usepackage{bm}
\usepackage[all]{hypcap}
\usepackage{colortbl}
\usepackage{booktabs}
\usepackage{braket}
\usepackage{mathrsfs}
\usepackage{multirow}
\usepackage{upgreek}
\usepackage{textcomp}
\usepackage{soul}
\usepackage{dsfont}
\usepackage[T1]{fontenc}
\usepackage{mathtools}
\usepackage{float}

\usepackage{lineno}

\usepackage{array}
\newcommand{\PreserveBackslash}[1]{\let\temp=\\#1\let\\=\temp}
\newcolumntype{C}[1]{>{\PreserveBackslash\centering}p{#1}}
\newcolumntype{R}[1]{>{\PreserveBackslash\raggedleft}p{#1}}
\newcolumntype{L}[1]{>{\PreserveBackslash\raggedright}p{#1}}

\newcommand{\degree}{^\circ}
\newcommand{\dupp}{\delta^\mathrm{upp}}
\newcommand{\dlow}{\delta^\mathrm{low}}
\newcommand{\freq}{\operatorname{freq}}
\newcommand{\ecomAT}{\epsilon_{\mathrm{com}}^{\mathrm{AT}}}

\newcommand{\vect}[1]{\mathbf{#1}}

\newcommand{\ustc}{
	\affiliation{Hefei National Research Center for Physical Sciences at the Microscale and School of Physical Sciences, University of Science and Technology of China, Hefei, China}
	\affiliation{CAS Center for Excellence in Quantum Information and Quantum Physics, University of Science and Technology of China, Hefei, Anhui, China}
	\affiliation{Hefei National Laboratory, University of Science and Technology of China, Hefei, China}
}
\newcommand{\ustcc}{
	\affiliation{Hefei National Research Center for Physical Sciences at the Microscale and School of Physical Sciences, University of Science and Technology of China, Hefei, China}
	\affiliation{CAS Center for Excellence in Quantum Information and Quantum Physics, University of Science and Technology of China, Hefei, Anhui, China}
}
\newcommand{\jinan}{
	\affiliation{Jinan Institute of Quantum Technology, Jinan, China}
}
\newcommand{\jinann}{
	\affiliation{Hefei National Laboratory, University of Science and Technology of China, Hefei, China}
	\affiliation{Jinan Institute of Quantum Technology, Jinan, China}
}
\newcommand{\Isreal}{
	\affiliation{The Center for Quantum Science and Technology, Department of Physics of Complex Systems, Weizmann Institute of Science, Rehovot, Israel}}

\newcommand{\Canada}{
	\affiliation{Institute for Quantum Computing and Department of Physics and Astronomy, University of Waterloo, Waterloo, Ontario, Canada.}}

\newcommand{\Singapore}{
	\affiliation{Department of Physics, National University of Singapore, Singapore, Singapore.}}

\newcommand{\coauthors}{
	\thanks{These authors contributed equally to this work.}
}
\begin{document}

\title{Device-independent quantum key distribution over\\100 km with single atoms}
\author{Bo-Wei Lu}\coauthors\ustc
\author{Chao-Wei Yang}\coauthors\ustc
\author{Run-Qi Wang}\coauthors\ustc
\author{Bo-Feng Gao}\jinan
\author{Yi-Zheng Zhen}\ustcc
\author{Zhen-Gang Wang}\ustc
\author{Jia-Kai Shi}\ustc
\author{Zhong-Qi Ren}\ustc
\author{Thomas A. Hahn}\Isreal
\author{Ernest Y.-Z. Tan}\Canada\Singapore
\author{Xiu-Ping Xie}\jinann
\author{Ming-Yang Zheng}\jinann
\author{Xiao Jiang}\ustc
\author{Jun Zhang}\ustc
\author{Feihu Xu}\ustc
\author{Qiang Zhang}\ustc\jinan
\author{Xiao-Hui Bao}\email{xhbao@ustc.edu.cn}\ustc
\author{Jian-Wei Pan}\email{pan@ustc.edu.cn}\ustc

\begin{abstract}
	\normalsize
	Device-independent quantum key distribution (DI-QKD) is a key application of the quantum internet. We report the realization of DI-QKD between two single-atom nodes linked by 100-kilometer (km) fibers. To improve the entangling rate, single-photon interference is leveraged for entanglement heralding, and quantum frequency conversion is used to reduce fiber loss. A tailored Rydberg-based emission scheme suppresses the photon recoil effect on the atom without introducing noise. We achieved high-fidelity atom-atom entanglement and positive asymptotic key rates for fiber lengths up to 100~km. At 11~km, 1.2~million heralded Bell pairs were prepared over 624~hours, yielding an estimated extractable finite-size secure key rate of 0.112~bits per event against general attacks. Our results close the gap between proof-of-principle quantum network experiments and real-world applications.
\end{abstract}

\maketitle
\section*{Introduction}

Quantum key distribution (QKD) is one of the most successful applications of quantum information science. In the early stage of the quantum internet~\cite{wehner2018quantum}, QKD primarily relies on ``prepare-and-measure'' schemes, which may suffer from practical loopholes due to imperfect devices~\cite{xu_secure_2020}. At a more advanced stage of the quantum internet, once heralded entanglement becomes available, device-independent QKD~\cite{pironio2009device,arnon2018practical,TSB+22} can be performed, which provides a theoretically maximal level of cryptographic security permitted by quantum mechanics. Unlike conventional QKD, DI-QKD achieves security from the observed violation of a Bell inequality~\cite{Ekert1991Quantum}, without requiring trust in the quantum devices' internal workings.

\begin{figure*}
	\includegraphics[width=0.9\linewidth]{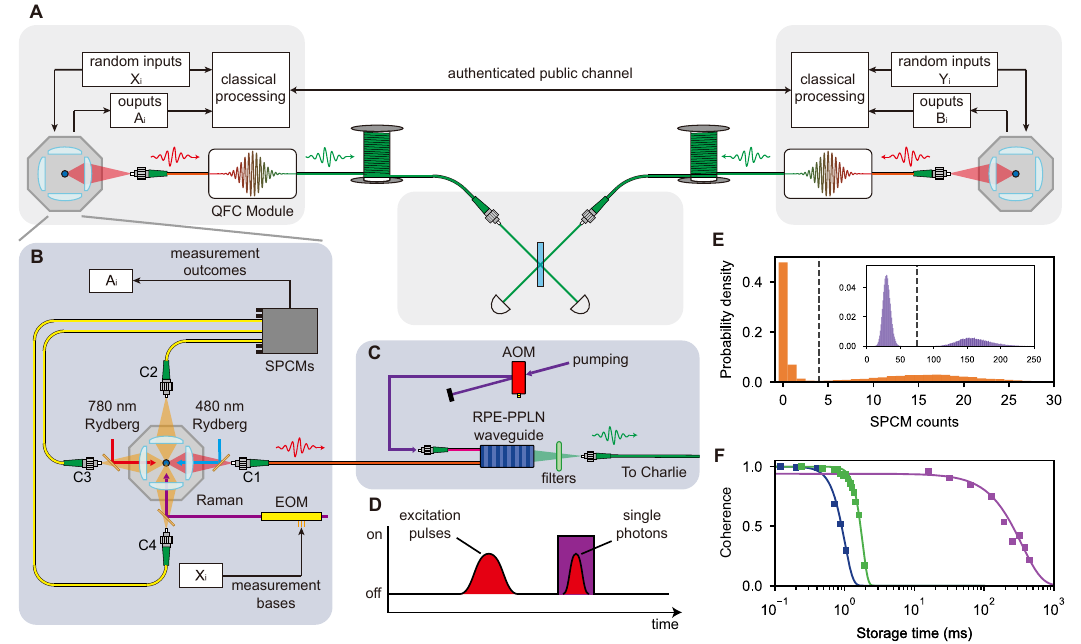}
	\caption{\textbf{Experimental architecture for device-independent QKD with neutral-atom quantum memories.} 
		(\textbf{A}) Overview of DI-QKD system. Two single-atom nodes (Alice and Bob) emit photons that are converted to the telecom band and interfered at Charlie for heralded entanglement. Random inputs $X_i$, $Y_i$ set measurement bases, and outputs $A_i$, $B_i$ are recorded. An authenticated public classical channel connects Alice and Bob for post-processing. 
		(\textbf{B}) Single-atom node schematic. The hyperfine qubits are manipulated via Raman pulses with electro-optic modulators (EOM) setting the measurement bases. A pair of Rydberg lasers drive the RSPE process for single-photon generation. Four in-vacuo lenses (NA = 0.52) collect fluorescence, with C2-C4 coupled to SPCMs for fluorescence detection and C1 coupled into a single-mode fiber to the QFC module. 
		(\textbf{C}) QFC module. The 780\,nm photons are converted to 1315\,nm in an RPE-PPLN waveguide, driven by a 1917\,nm pump and followed by spectral filtering. An AOM gates the pump to suppress noise.
		(\textbf{D}) Temporal filtering. The pump is switched on only during photon arrival to suppress excitation-pulse noise.
		(\textbf{E}) Fluorescence detection. Histogram of SPCM counts with 35\,$\mu$s exposure; dashed line marks the discrimination threshold (4 counts). The inset shows photon-counting statistics from 5-ms weak imaging with single-lens collection (C4), yielding a state discrimination fidelity above 99.9\%.
		(\textbf{F}) Long-lived quantum memories. Qubit coherence versus storage time for Hahn-echo (blue), XY-4 (green), and transferring qubits into clock basis and then applying XY-8-32 sequence (purple).
		}
	\label{fig:setup}
\end{figure*}

Realizing DI-QKD requires the creation of remote entanglement with sufficiently high fidelity together with high detection efficiency above a threshold required for violating Bell inequalities without detection loopholes. These demanding requirements have so far restricted DI-QKD demonstrations to laboratory settings with very short distances~\cite{nadlinger2022experimental,zhang2022device,liu2022toward}. With trapped ions~\cite{nadlinger2022experimental}, finite-key security was achieved over 3.5~meters in fiber. With single atoms~\cite{zhang2022device} or entangled photons\cite{liu2022toward}, the feasibility of achieving positive asymptotic key rates over several hundreds meters was demonstrated. Although long-distance photonic DI-QKD protocols~\cite{PhysRevLett.105.070501} have been proposed, they still face substantial practical challenges in achieving the necessary fidelity and detection efficiency over extended distances.

Meanwhile, progress has been made in extending the entangling distance for matter qubits~\cite{yu2020entanglement,van2022entangling,liu2024creation,knaut2024entanglement,stolk2024metropolitan}. Quantum frequency conversion (QFC)~\cite{zaske2012visible,de2012quantum} has enabled photons emitted at atomic resonance to be converted to the low-loss telecom band, facilitating entanglement distribution over long-distance fibers. Additionally, single-photon interference (SPI)-based entanglement generation~\cite{cabrillo1999creation} has contributed to extending this distance~\cite{yu2020entanglement,liu2024creation,stolk2024metropolitan}. Nevertheless, these advances have been hampered by low fidelities and entangling rates, failing to meet the requirements for DI-QKD.

We demonstrate DI-QKD by distributing high-fidelity heralded entanglement between two $^{87}$Rb atoms over spooled fibers up to 100~km. Remote entanglement is generated through the SPI scheme, which mitigates photon loss and enables efficient long-distance entanglement distribution. By using a tailored Rydberg-based single-photon emission process, the photon recoil effect in SPI is strongly suppressed without introducing noise. Fiber attenuation is further reduced by down-converting the emitted 780-nm photons to a 1.3-$\mu$m telecom wavelength. Together, these advances enable long-distance DI-QKD. We realized DI-QKD over 11~km with finite-key security and achieved a positive asymptotic key rate over 100~km.

\begin{figure*}
	\includegraphics[width=0.75\linewidth]{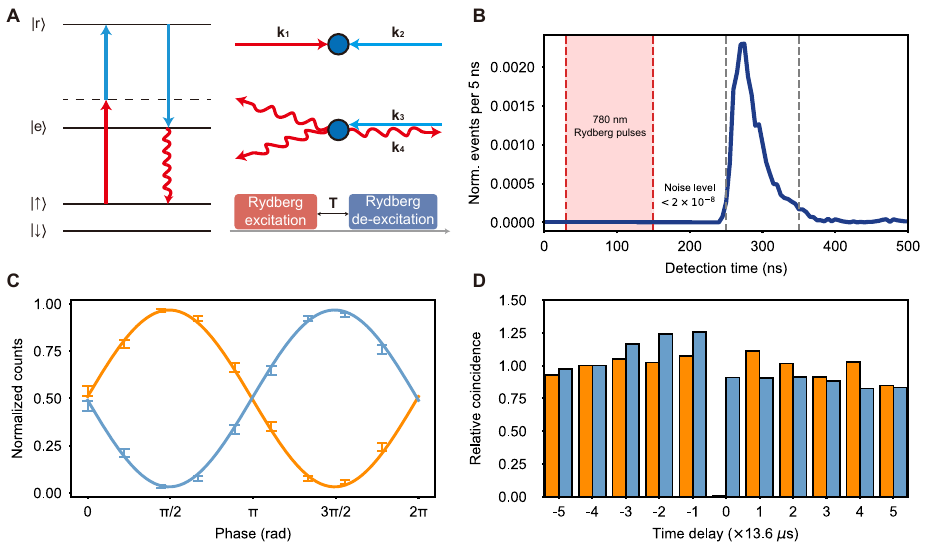}
	\caption{\textbf{Single-photon generation and characterization.} 
			(\textbf{A}) Rydberg-based single photon emission (RSPE) scheme. The qubit state $\ket{\uparrow}$ is coherently excited to a long-lived Rydberg state $\ket{r}$ via a two-photon transition ($k_1$, $k_2$). After a controllable delay $T$, the atom is de-excited to the excited state $\ket{e}$ ($k_3$), from which spontaneous decay produces a single photon $(k_4)$.  
			(\textbf{B}) Temporal profile of detected photons. The 780\,nm Rydberg excitation pulses (red region) are well separated from the single-photon signal (blue peak); The noise level is suppressed to below $2\times10^{-8}$ in normalized units.
			(\textbf{C}) Single-photon interference fringes obtained by scanning the phase of the interferometer and detecting the normalized single-photon counts from two output channels. Blue and orange points correspond to two SNSPD channels, respectively. Solid curves represent sinusoidal fits to the measured data. Error bars represent one standard deviation assuming Poissonian statistics.
			(\textbf{D}) Hong-Ou-Mandel (HOM) interference of independently emitted photons. Relative coincidence counts for parallel (orange) and perpendicular (blue) polarizations are shown as a function of detection time delay.   
	}
	\label{fig:RSPE}
\end{figure*}
\section*{Experimental architecture} 
Our experimental platform, which was designed to address the challenges of heralded entanglement distributionover long distances, consists of two independent quantum network nodes (Alice and Bob). These nodes are connected by optical fibers to a central Bell state measurement (BSM) node, Charlie. All three nodes are housed within the same laboratory.

At each node, a single $^{87}$Rb atom is trapped in a pair of crossed optical tweezers and serves as a long-lived quantum memory. The atoms are confined at the common focus of four in-vacuum aspheric lenses (numerical aperture (NA) = 0.52), which efficiently collect atomic fluorescence (Fig.~\ref{fig:setup}B). This high-performance atom-photon interface enables fast fluorescence detection within 35~$\mu$s, with a discrimination fidelity exceeding 99.5\% (Fig.~\ref{fig:setup}E). Although the present implementation does not close the locality loophole, such fast, high-fidelity readout constitutes an essential prerequisite for future Bell tests that could achieve this benchmark.

The atomic qubit is encoded in a pair of hyperfine ground states, $\ket{\uparrow} \equiv \ket{F = 2, m_F = -2}$ and $\ket{\downarrow} \equiv \ket{F = 1, m_F = -1}$. High-fidelity single-qubit rotations between these states are implemented via two-photon Raman transitions~\cite{levine2022dispersive}. Remote entanglement between the two atomic memories is generated using the SPI scheme. Each trial begins with optical pumping to $\ket{\downarrow}$, followed by a coherent rotation into an unbalanced superposition $\sqrt{\alpha}\ket{\uparrow}+\sqrt{1-\alpha}\ket{\downarrow}$, where $\alpha$ denotes the excitation probability. The $\ket{\uparrow}$ state is then selectively excited to an excited state $\ket{e}$, from which spontaneous decay produces a single photon, entangling the atomic qubit with the photon's presence. 

The emitted photons from Alice and Bob are directed to Charlie, where they interfere on a balanced beam splitter. A single detection event at either output heralds entanglement between the two remote atomic memories. Due to channel loss, single-click events can occur when both nodes emit photons. Consequently, a detector click heralds a mixed state:
\begin{equation}
	\rho_{AB} 
	= \alpha\ket{\uparrow\uparrow}\bra{\uparrow\uparrow}+(1-\alpha)\ket{\psi^{\pm}}\bra{\psi^{\pm}},
	\label{eq:memory-memory-entanglement}
\end{equation}
where $\ket{\psi^{\pm}} \equiv (\ket{\uparrow\downarrow}\pm e^{i\delta\phi}\ket{\downarrow\uparrow})/\sqrt{2}$ denotes the maximal entangled states, and the sign depends on which output the photon is detected in. The relative phase $\delta\phi$, determined by the optical path difference between the interferometer arms, is actively stabilized throughout the experiment (see Ref.~\cite{SM} for details). Compared with the two-photon interference (TPI) approach~\cite{van2022entangling,krutyanskiy2023entanglement}, SPI requires only a single detector click to herald entanglement, substantially enhancing the generation rate. This advantage is essential for meeting the stringent statistical requirements of practical DI-QKD.

Heralded entanglement between matter qubits based on the SPI scheme has been constrained by relatively low fidelity, even at short distances~\cite{ruskuc2025multiplexed,humphreys2018deterministic,yang2025entangling}. For weakly bound neutral atoms, achieving high-fidelity entanglement via SPI is even more challenging -- the recoil momentum of the photon alters the atomic motional state, revealing the photon's ``which-path'' information, thereby reducing the fidelity of the generated entanglement~\cite{slodivcka2013atom}. To suppress the photon recoil effect, the excitation pulses need to co-propagate with the single photons emitted by the atoms, minimizing the net recoil momentum. Under resonant excitation, the emitted single photons are overwhelmed by noise from excitation pulses, reducing the fidelity of the entanglement. 

To address this issue, we use a Rydberg-based single-photon emission (RSPE) protocol for atom-photon entanglement generation, which is commonly employed to suppress multiple excitations in atomic ensembles in Duan-Lukin-Cirac-Zoller protocol~\cite{yang2025entangling,dudin2012strongly}, but here it is used for temporal filtering (Fig.~\ref{fig:RSPE}A). In this scheme, the $\ket{\uparrow}$ state is selectively excited to a Rydberg state $\ket{r}$ with a lifetime of tens of microseconds. After a controllable delay, $\ket{r}$ is coherently transferred to $\ket{e}$, from which spontaneous decay produces the emitted photons. Since only the photons co-propagating with the 780~nm Rydberg pulses are coupled into the collection mode, the relevant momentum transfer can be referenced to this mode. The net momentum imparted to the atom in the RSPE process is given by 
$\Delta \mathbf{k} = \mathbf{k}_1 + \mathbf{k}_2 - \mathbf{k}_3 - \mathbf{k}_4$, where $\mathbf{k}_i$ are the wave vectors of the four interacting optical fields (see Fig.~\ref{fig:RSPE}A). By employing a collinear geometry, we minimize $\Delta \mathbf{k}$ while also temporally separating photon generation from the excitation pulse, thereby suppressing recoil-induced decoherence without introducing additional noise. The crossed optical tweezers further localize the atom, increasing the Debye-Waller factor and further mitigating recoil effects. Thanks to this tailored excitation scheme, the cycling transition $\ket{\uparrow}\leftrightarrow\ket{e}\equiv\ket{F'=3, m_{F}'=-3}$ on the D$_2$ line can be used to generate atom-photon entanglement, avoiding state leakage and maximize the transition branching ratio, hense preserve purity and maximize the efficiency of the photon.

As a benchmark for photon coherence, we performed an experiment analogous to Young's double-slit~\cite{fedoseev2025coherent,zhang2025Tunable}, where the atoms at Alice and Bob act as two effective slits whose emissions interfere at the central beam splitter. By controlling the pulse area of the Rydberg excitation beams, we set the excitation probability to $p_r = 0.02$, thereby preparing the emitted photons in a superposition of the vacuum and single-photon states $\sqrt{1-p_r}\ket{0}+\text{e}^{i\varphi}\sqrt{p_r}\ket{1}$. The relative phase $\varphi$ is tuned by varying the interferometer's locking point with an electro-optic modulator (EOM). We observed clear interference fringes (Fig.~\ref{fig:RSPE}C) with a visibility of $0.93 \pm 0.01$, which is limited primarily by residual photon recoil. These results confirm that both photon coherence and interferometric stability meet the requirements for high-fidelity BSM in the SPI protocol. Further analysis of photon recoil in the SPI scheme is provided in Ref.~\cite{SM}.

In the RSPE process, unwanted re-excitations are strongly suppressed because the $\ket{\uparrow}\to\ket{r}$ transition is driven via a far-detuned two-photon Raman process, which reduces off-resonant scattering and thereby suppresses re-excitation -- a common source of error in resonant excitation schemes. To benchmark the indistinguishability of the photons generated via the RSPE process, we performed Hong-Ou-Mandel (HOM) interference measurements between independently emitted  780~nm photons by Alice and Bob without spectral filtering (Fig.~\ref{fig:RSPE}D). Within a 100~ns detection window, where over 96\% of the photons are detected, the measured visibility of $0.991\pm0.01$ demonstrates near-perfect indistinguishability, which is crucial for achieving high-fidelity atom-atom entanglement.

\begin{figure*}
	\includegraphics[width=0.9\linewidth]{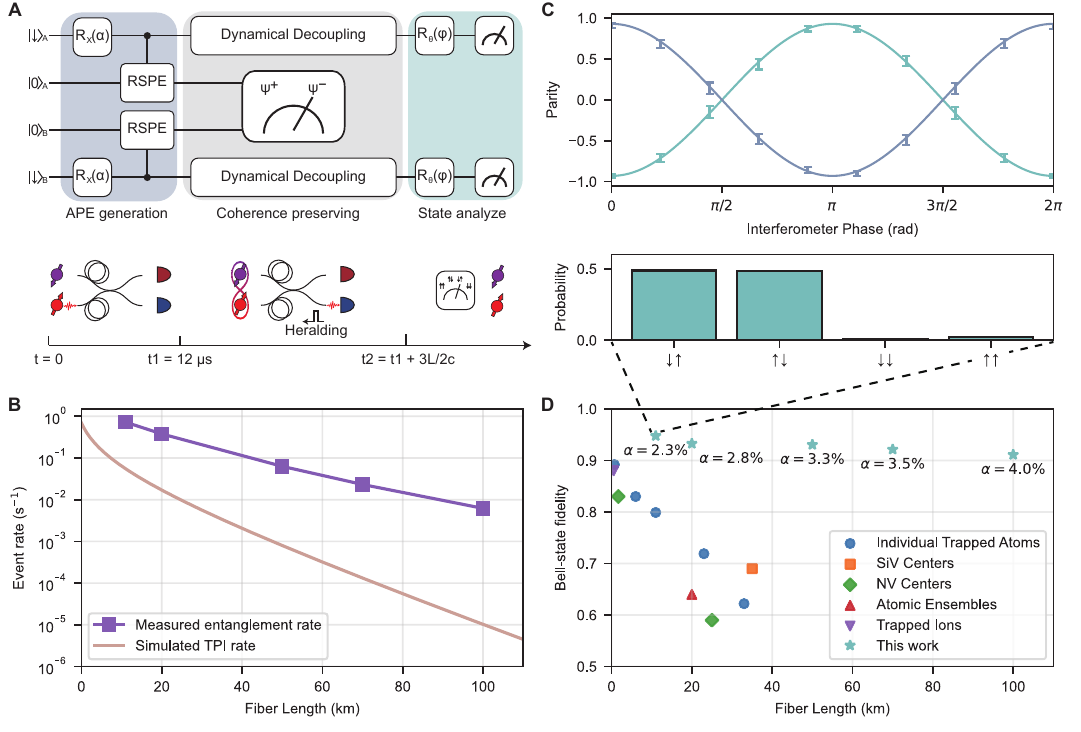}
	\caption{\textbf{Heralded entanglement over long-distance telecom fiber.} 
			(\textbf{A}) Experimental sequence for atom-atom entanglement. Atom–photon entanglement (APE) is first generated at each node, and the photons interfere at Charlie to herald entanglement. Memory coherence is preserved using dynamical decoupling. The entangled state is analyzed by applying single-qubit rotations $R_\theta(\varphi)$ followed by push-out detection.
			(\textbf{B}) Heralded entanglement event rate versus total fiber length $L$. Purple squares denote measured rates with SPI Scheme, and brown line denotes the expected rate for two-photon interference scheme at the same repetition rate, which scales with $0.5\eta_A\eta_B$ and falls more steeply with distance. 
			(\textbf{C}) Entanglement characterization at 11 km. Top: parity oscillations for the $\ket{\psi^{+}}$ (teal) and $\ket{\psi^{-}}$ (blue-gray) states versus interferometer phase, with sinusoidal fits (solid curves). 
			Bottom: measured populations in the $\hat{Z}\hat{Z}$ basis, with statistical uncertainties below $10^{-3}$.
			(\textbf{D}) Entanglement fidelity versus $L$. 
			Stars: this work (labels indicate the optimized excitation probability $\alpha$ at each $L$); points: prior demonstrations (individual atoms~\cite{van2022entangling,zhang2022device}, NV/SIV centers~\cite{knaut2024entanglement,stolk2024metropolitan,hensen2015loophole}, trapped ions~\cite{krutyanskiy2023entanglement}, atomic ensembles~\cite{liu2024creation}). 
			Fidelities remain $> 0.9$ up to 100~km, satisfying DI-QKD requirements. }
	\label{fig:entanglement}
\end{figure*}
\section*{Heralded Atom-Atom Entanglement}
\begin{figure*}
	\includegraphics[width=0.9\linewidth]{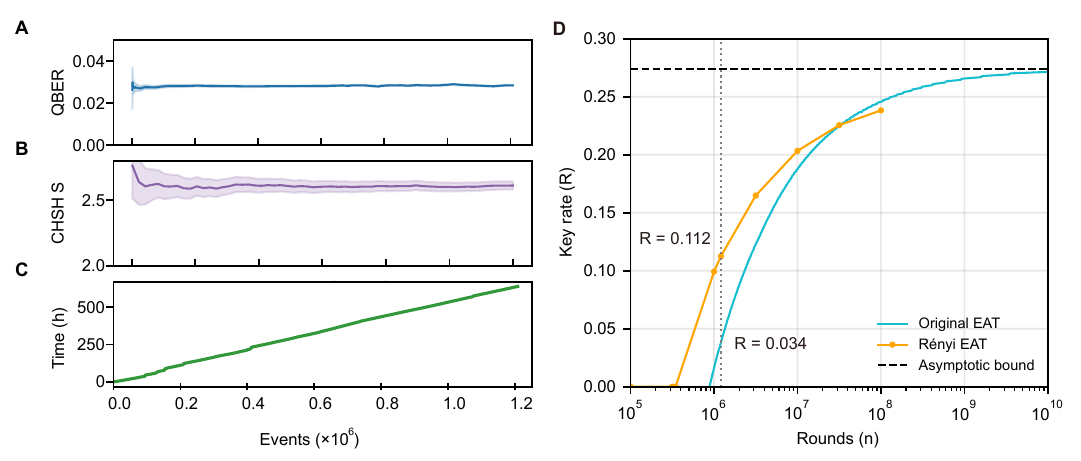}
	\caption{\textbf{DI-QKD over 11\,km with finite-size security.} 
		\textbf{A–C}, Experimental performance metrics.   
		\textbf{A}, Quantum bit error rate (QBER); \textbf{B}, CHSH Bell parameter $S$ versus accumulated events, with shaded bands indicating one standard deviation.   
		\textbf{C}, Wall-clock time to accumulate events.  
		\textbf{D}, Finite-size secure key rate $R$ versus the number of rounds $n$ with a soundness error of $\epsilon_{\mathrm{snd}} = 10^{-5}$ and optimized Bell-test fractions ($\gamma_A = 0.26$, $\gamma_B = 0.13$). evaluated using two methods (original EAT and R\'{e}nyi EAT). Cyan: original EAT; orange: R\'{e}nyi EAT; dashed: asymptotic bound. At $n = 1.208\times10^{6}$ (dotted line), the R\'{e}nyi EAT yields $R = 0.112$ bits per event (0.06 bit/s), while the original EAT gives $R = 0.034$ bits per event.}
	\label{fig:finite_size}
\end{figure*}
The emitted 780~nm photons are down-converted to the telecom O-band (1315~nm) with 1917~nm pumping using reverse proton-exchange periodically poled lithium niobate (PPLN) waveguides~\cite{SM}. This reduces fiber attenuation from $\sim$3~dB/km at 780 nm to $\sim$0.32~dB/km at 1315~nm. The QFC module, including wavelength conversion, spectral filtering, and fiber coupling, achieves an overall efficiency of 47\%. During the excitation sequence, the QFC pumping laser is gated by an acousto-optic modulator (AOM) to avoid detector dead time caused by excitation pulses (Fig.~\ref{fig:setup}D). An additional 300~MHz filter cavity is inserted after the BSM to suppress out-of-band noise. Together with the temporal filtering mentioned above, we measured noise count rates below 0.5 Hz and a signal-to-noise ratio (SNR) exceeding 30,000:1 with $\alpha = 1$ immediately after the QFC module within a 100~ns window (Fig.~\ref{fig:RSPE}B), compared to the SNR of 20:1 in our previous work~\cite{yu2020entanglement}. This ensures high SNR even over 100-km fiber links.

We then evaluate the fidelity of atom-atom entanglement distributed across long fibers. The entanglement is created in an "event-ready" fashion, where both nodes delay their measurements by $3L/2c$ after photon emission, with $L$ denoting the total fiber length to Charlie and c representing the speed of light in vacuum. This waiting time accounts for the single-photon propagation to Charlie and the subsequent return of the classical heralding signal to each node (Fig.~\ref{fig:entanglement}A). The event-ready entanglement reduces the need for atom reloading, thereby enhancing the entangling rate. To preserve qubit coherence during the delay, we apply dynamical decoupling (DD) sequences to each atomic memory, as shown in Fig.~\ref{fig:setup}F. The coherence can be further prolonged to over 300 ms by transferring the qubit into magnetically insensitive clock states~\cite{SM}.

\begin{figure}
	\includegraphics[]{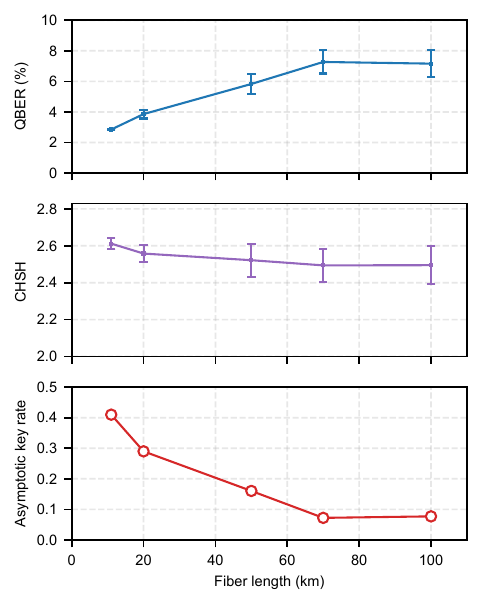}
	\caption{
		\textbf{Device-independent QKD performance over metropolitan-scale fiber.}
		\textbf{A}, Quantum bit error rate (QBER) versus fiber length, increasing from $2.85\%$ at 11\,km to $7.16\%$ at 100\,km.  
		\textbf{B}, CHSH violation S as a function of fiber length, remaining well above the classical bound of 2 for all distances tested, thus confirming sustained nonlocal correlations. 
		\textbf{C}, Asymptotic secret key rate with $\gamma = 10^{-3}$, showing secure key generation maintained up to 100\,km. Error bars represent one standard deviation from Poissonian counting statistics.
	}
	\label{fig:DIQKD_performance}
\end{figure}
We characterize the entanglement fidelity for fiber lengths of 11, 20, 50, 70, and 100~km. Upon heralding by a single-photon detection, the atomic hyperfine qubits are measured using the push-out method, with the measurement basis set by a Raman pulse. Since the measurement events in this experiment are not space-like separated, fluorescence detection with a single aspheric lens is employed with 5~ms imaging (inset of Fig.~\ref{fig:setup}E) in the following experiments. The fidelity is evaluated by performing projective measurements of the atomic qubits in both the $\hat{X}\hat{X}$ and $\hat{Z}\hat{Z}$ bases, which together provide a direct estimate of the coherence and population components of the entangled state. It is calculated as~\cite{guhne2009}
\begin{equation}
	\mathcal{F} = \frac{1}{4}\left[1 + \mathcal{V}_{ZZ} + 2\mathcal{V}_{XX}\right],
	\label{eq:fidelity}
\end{equation}
where $\mathcal{V}_{ZZ}$ and $\mathcal{V}_{XX}$ are the measured visibilities in the respective bases. For each fiber length, the excitation probability $\alpha$ is optimized to maximize fidelity, balancing errors from higher-order excitations at short links (favoring smaller $\alpha$) against background noise at longer links (favoring larger $\alpha$). At each distance, the optimized $\alpha$ is shown in Fig.~\ref{fig:entanglement}D. Representative results at 11 km, obtained using a Hahn-echo sequence with a total decoupling time of 55~$\mu$s, are shown in Fig.~\ref{fig:entanglement}b. For longer fiber lengths, we employ XY-4 dynamical decoupling sequences extend memory coherence. The measured fidelities are $0.947\pm0.005$ (11~km), $0.933\pm0.006$ (20~km), $0.931\pm0.008$ (50~km), $0.921\pm0.009$ (70~km), and $0.911\pm0.010$ (100~km), each averaged over $\ket{\psi^+}$ and $\ket{\psi^-}$. The fidelity remains above 0.9 across all tested fiber lengths up to 100~km (Fig.~\ref{fig:entanglement}D), confirming the robustness of the protocol under practical conditions and demonstrating state-of-the-art performance.

Another key performance metric is the entanglement event rate, which directly determines the secure key generation rate in DI-QKD. The rate is given by the product of the success probability $p_s$ per trial and the experimental repetition rate $R$. As the fiber length increases, both the fiber transmission decreases and the communication overhead increases, leading to a reduction in the event rate. In the SPI scheme, the success probability is
\begin{equation}
	p_s = \alpha_A\eta_A+\alpha_B\eta_B,
	\label{eq:success_prob}
\end{equation}
where $\alpha_{A(B)}$ denotes the excitation probability at Alice (Bob), and $\eta_{A(B)}$ represents the overall photon transmission and detection efficiency from each node to Charlie. $\alpha$ and $\eta$ are well balanced in the experiment. Compared with TPI schemes, where $p_s = 0.5\eta_A \eta_B$, the linear dependence in SPI significantly boosts the entanglement generation rate, particularly over long distances (Fig.~\ref{fig:entanglement}C). Further details on the entanglement event rate and the loss budget are provided in Ref.~\cite{SM}.

\section*{Implementation of DI-QKD}

With the ability to reliably distribute high-fidelity entanglement over metropolitan-scale fiber links, we proceed to demonstrate the implementation of DI-QKD (Fig.~\ref{fig:setup}.A), in which security is certified by the violation of a Bell inequality. In each trial, a successful heralding event triggers real-time random basis selection at Alice ($\{\hat{Z},(\hat{Z}+\hat{X})/\sqrt{2},(\hat{Z}-\hat{X})/\sqrt{2}\}$ for $X_i\in\{0,1,2\}$) and Bob ($\{\hat{Z},\hat{X}\}$ for $Y_i\in\{0,1\}$) using independent quantum random number generators (QRNGs). No classical negotiation of measurement settings is performed during the measurement phase. A subset of the events ($X_i \in \{1,2\}$ and $Y_i \in \{0,1\}$, chosen with probabilities $\gamma_A$ and $\gamma_B$, respectively) is used for Bell tests evaluating the CHSH violation $S$, another subset ($X_i = 0$ and $Y_i = 0$) is used for key generation in the $\hat{Z}\hat{Z}$ basis, and events with other basis choices are discarded and kept private. No measurement outcomes of these discarded rounds is shared between Alice and Bob, ensuring that they do not compromise the privacy of the generated key. A detailed protocol is described in Ref.~\cite{SM}.

In the implementation of DI-QKD, ensuring the privacy of the measurement outcomes and bases is crucial. To prevent 780 nm fluorescence leakage during atomic state measurement, the QFC pump laser is synchronized with the entanglement generation trial. The pump laser is turned on for 200 ns in each trial and kept off for the rest of the time. Unconverted 780 nm photons, as well as scattered Raman pulses which determine the measurement basis, are attenuated by the waveguide and subsequent spectral filters with total transmission below $10^{-7}$.

We first analyze the finite-size performance of the system at a total fiber length of 11 km, a distance that mimics metropolitan-scale QKD requirements with relatively high heralding rates. The fractions of trials allocated to Bell tests are optimized to $\gamma_A = 0.13$ and $\gamma_B = 0.26$. Phase stabilization of the SPI interferometer is maintained continuously via an FPGA-based feedback loop, requiring no manual intervention. Both nodes operate in independent temperature-controlled enclosures with peak-to-valley fluctuations below 0.03 $\degree$C, providing passive stability over multi-day runs.

Over 624~h of operation, we collected $1.208\times10^6$ heralded entanglement events (Fig.~\ref{fig:finite_size}C), corresponding to a mean event rate of 0.53~s$^{-1}$. The run included intermittent pauses caused by laser failures and one recalibration of the photon coupling efficiency, but otherwise remains stable. Compared to the 0.72~s$^{-1}$ observed in Fig.~\ref{fig:entanglement}B, the slight reduction is attributed to long-term drifts in efficiency. From this data, we measured a CHSH violation of $S = 2.612 \pm 0.031$ and a quantum bit error rate (QBER) of $Q = 0.0285 \pm 0.0002$, both indicating high-quality entanglement suitable for DI-QKD (Fig.~\ref{fig:finite_size}A,B). Using a recently developed entropy accumulation theorem method based on R\'{e}nyi entropy~\cite{arq2024generalized}, we obtain an estimated secure key rate of 0.112~bits per event against general attacks with a soundness error of $\epsilon_{\mathrm{snd}} = 10^{-5}$, compared to the asymptotic value of 0.275 (Fig.~\ref{fig:finite_size}D). With this finite-size key rate, the extractable secure key is estimated to be 135.3 kbits, with an average key generation rate of 0.06 bit/s. Even with a tightened soundness error of $\epsilon_{\mathrm{snd}} = 10^{-15}$, the rate remains positive at 0.075~bits per event. Note that the original EAT method~\cite{arnon2018practical} still gives a positive finite-size key rate of 0.034~bits per event with $\epsilon_{\mathrm{snd}} = 10^{-5}$.
 
As the number of rounds increases, the R\'{e}nyi EAT approach taken in this work performs suboptimally because it does not capture the accumulation of entropy during test rounds; this can be improved upon by either using a more sophisticated R\'{e}nyi EAT toolkit from~\cite{arq2024generalized} or by decreasing the test-round fraction as the proposed number of rounds increases. These results demonstrate, for the first time, the feasibility of finite-size DI-QKD at the 10~km scale.

We finally evaluate the asymptotic performance of the system at 11, 20, 50, 70, and 100~km (Fig.~\ref{fig:asymptotic}). For all tested links, we observe CHSH violations and $\hat{Z}\hat{Z}$-basis correlations that yield positive key rates with $\gamma = 10^{-3}$. The p-value analysis of the observed CHSH violation indicates that the probability of the data being consistent with classical correlations under the null hypothesis is below $5\times10^{-6}$ for all tested fiber lengths~\cite{SM}. This serves as an upper benchmark for the ultimate capability of the memory-assisted DI-QKD system. The analysis shows that secure key generation remains feasible across the entire range, with the key rate decreasing as fiber length increases due to reduced heralding probability and lower SNR.

\section*{Discussion and outlook}
We have successfully generated heralded entanglement over metropolitan distances with high fidelity between two single-atom based quantum nodes, and realized DI-QKD. The Rydberg-based photon emission scheme allows us to minimize photon recoil effect without introducing additional noise, thereby preserving the high purity of the emitted photons. The utilization of multiple in-vacuum lenses enables fast atomic-state detection. The use of single-photon interference scheme for remote entanglement heralding enables us to obtain an metropolitan entangling speed that is orders of magnitude higher than the two-photon based schemes used in previous DI-QKD experiments~\cite{nadlinger2022experimental,zhang2022device}. 

The demonstration of device-independent QKD at the metropolitan scale helps close the gap between proof-of-principle quantum network experiments and real-world applications. Beyond DI-QKD, the architecture demonstrated here offers a versatile platform for device-independent quantum random number generation (DI-QRNG)~\cite{MayersYao1998Quantum}, self-testing of quantum devices~\cite{vsupic2020self}, and fundamental test of quantum mechanics~\cite{hensen2015loophole,rosenfeld2017event,giustina2015significant,shalm2015strong}. Moreover, the high-fidelity entanglement we demonstrate can serve not only as a valuable resource for quantum network applications, but also as a fundamental building block for scaling up quantum networks.

Further extension of the entangling distance can be achieved by reducing channel attenuation, for example, through quantum frequency conversion to low-loss 1.5~$\mu$m wavelengths~\cite{ikuta2011wide,van2020long} and by using hollow-core fibers~\cite{petrovich2025broadband}. Our experimental setup is compatible with Rydberg-based tweezer arrays, enabling multiplexed quantum networking for parallel entanglement generation~\cite{hartung2024quantum,li2025parallelized}, supporting entanglement purification~\cite{Bennett96,pan_entanglement_2001,pan_experimental_2003,kalb2017entanglement} and deterministic entanglement swapping via Rydberg-based atom-atom gates~\cite{evered2023high}. With feasible improvements, the entangling rate over a 10 km link could exceed 10 kHz~\cite{SM}.
Together, our results illustrate the promise for a scalable approach for the development of quantum networks~\cite{azuma_quantum_2023}.

\section*{Acknowledgments}
\textbf{Funding:}
This research was supported by the Quantum Science and Technology-National Science and Technology Major Project (No. 2021ZD0301101, No. 2021ZD0301104, No. 2021ZD0300802, No. 2021ZD0300303), National Natural Science Foundation of China (No.~T2525008, No.~T2125010, No.~12274394, No.~62031024, No.~12475028, No.~12505018), National Key R\&D Program of China (No. 2020YFA0309804), and the Chinese Academy of Sciences. C.W.Y. acknowledges support from China Postdoctoral Science Foundation (No. BX20230105, No. 2023M730901). B.F.G. acknowledges support from the Natural Science Foundation of Shandong Province, China (No. ZR202211110166). Y.Z.Z. acknowledges support from Anhui Provincial Natural Science Foundation (No. 2308085MA26) and the Fundamental Research Funds for the Central Universities (No. WK9990000125). T.A.H. was supported by the Koshland Research Fund and by the Air Force Office of Scientific Research under the award number FA9550-22-1-0391. E.Y.Z.T. performed this research partly while at the National University of Singapore and partly while at the Institute for Quantum Computing (University of Waterloo), supported by Innovation, Science, and Economic Development Canada, as well as NSERC under the Discovery Grants Program, Grant No. 341495. F. Xu thanks the support by the New Cornerstone Science Foundation through the Xplorer Prize. The numerical calculations in this paper have been performed on the supercomputing system in the Supercomputing Center of University of Science and Technology of China.

\textbf{Author contributions:}
J.-W.P. conceived the research. X.-H.B. and J.-W.P. designed the experiment. B.-W.L., C.-W.Y., R.-Q.W., Z.-G.W. and J.-K.S. carried out the experiment and analyzed the data. B.-F.G., X.-P.X., M.-Y.Z. and Q.Z. contributed the QFC modules. Y.-Z.Z., Z.-Q.R., T.A.H., E.Y.-Z.T. and F.X. contributed analysis on secure key rates. X.J. contributed the electronic board for phase stabilization. J.Z. contributed the QRNGs. B.-W.L., C.-W.Y., R.-Q.W., X.-H.B. and J.-W.P. wrote the manuscript with input from all other authors. X.-H.B. and J.-W.P. supervised the whole project.

\textbf{Competing interests:}
There are no competing interests to declare.

\textbf{Data and materials availability:} The data underlying the figures are deposited at Zenodo~\cite{Zenodo}. All other data needed to evaluate the conclusions in the paper are present in the main text or the Supplementary Material.

\setcounter{figure}{0}
\setcounter{table}{0}
\setcounter{equation}{0}

\onecolumngrid

\global\long\def\theequation{S\arabic{equation}}
\global\long\def\thefigure{S\arabic{figure}}
\global\long\def\thetable{S\arabic{table}}
\renewcommand{\arraystretch}{0.6}

\newpage

\newcommand{\msection}[1]{\vspace{\baselineskip}{\centering \textbf{#1}\\}\vspace{0.5\baselineskip}}

\msection{SUPPLEMENTAL MATERIAL}

\subsection{Details on quantum node}
Each quantum network node is built around a titanium high-vacuum chamber housing a single $^{87}$Rb atom. Four aspheric lenses (Thorlabs, AL2550H, NA = 0.52), labeled AL1-AL4, are installed inside the chamber. AL1 and AL3 are aligned along the quantization axis, while the remaining lenses are arranged orthogonally. Their positions are carefully aligned prior to bake-out to minimize aberrations and maximize photon collection efficiency.

The atom is trapped at the common focus of two 852-nm optical tweezers, incident through AL2 and AL3, with waists of 1.50 $\mu$m and 1.61 $\mu$m and trap depths of 1.0 mK and 1.2 mK, respectively. During entanglement generation, the trap depth is adiabatically reduced to one third of its initial value, thereby increasing coherence time and mitigating parametric heating from repeated switching of the tweezers during the Rydberg sequences. The atom is cooled to approximately 15 $\mu$K via a 3D polarization-gradient cooling stage, with the temperature measured using the release-and-recapture method.

Single photons generated by the Rydberg-based single photon emission process are collected and collimated by AL1. They are then coupled into an AR-coated single-mode fiber (Corning HI-780) using an external aspheric lens (Thorlabs, AL50100H). A band-pass filter suppresses background light. The overall efficiency—including collection, transmission, and fiber coupling—is measured to be approximately 4\%, corresponding to a coupling efficiency of about 50\% from AL1 into the fiber. The interface for fluorescence detection consists of AL2-AL4, each coupled to multimode fibers. These fibers are linked to three single-photon counting modules (SPCMs).

\begin{figure}[htbp]
	\centering
	\includegraphics[width=.9\linewidth]{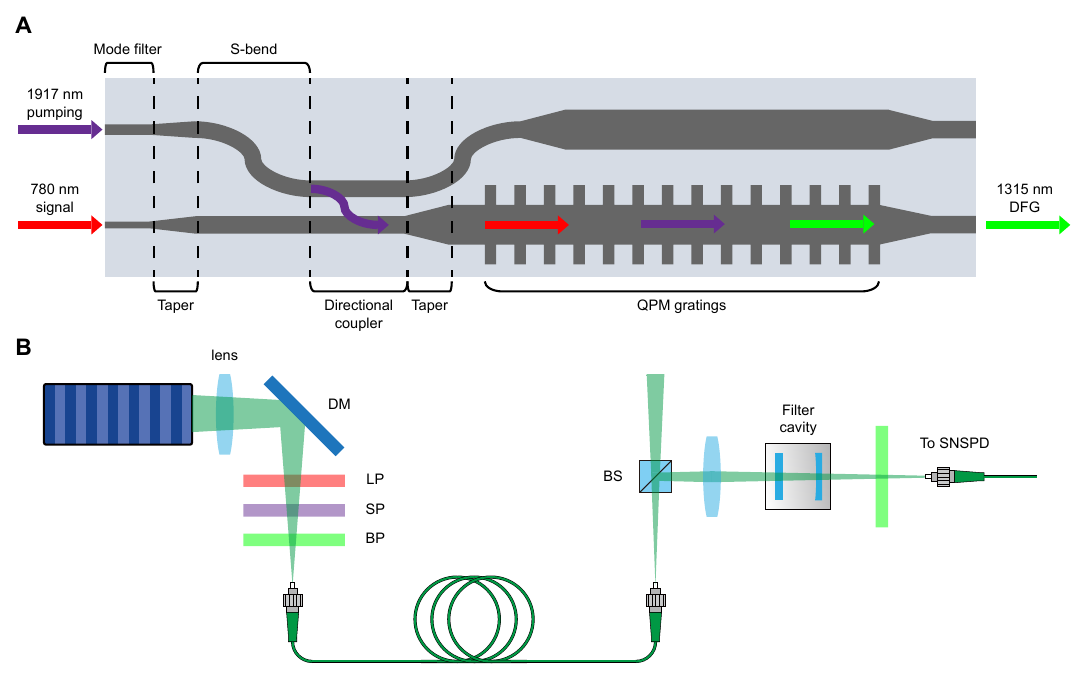}
	\caption{\textbf{Quantum frequency conversion setup.} 
		\textbf{A.} Structure of the RPE-PPLN waveguide, showing mode filters, directional coupler, tapers, and quasi-phase-matched (QPM) gratings for converting 780-nm signal photons and 1917-nm pump photons into 1315-nm telecom photons.  
		\textbf{B.} Spectral filtering of the converted photons using a dichroic mirror (DM), cascaded long-pass (LP), short-pass (SP), and band-pass (BP) filters, followed by a filter cavity before detection with SNSPD.}
	\label{fig:QFC1}
\end{figure}a

\subsection{Details on quantum frequency conversion module}
The QFC chip and noise filtering is shown in Fig.~\ref{fig:QFC1}. The quantum frequency conversion is implemented using reverse-proton-exchange (RPE) periodically poled lithium niobate (PPLN) waveguide chips, which convert 780-nm signal photons to the 1315-nm telecom O-band under pumping by a 1917-nm laser. The 1917-nm pump and 780-nm signal are combined via an on-chip directional coupler. At the chip input, the mode filter width is 2 $\mu$m for the 780-nm signal and 5.5 $\mu$m for the 1917-nm pump. The central spacing between the two mode filters is 127$\mu$m. For the directional coupler, the waveguide width is 5.5 $\mu$m, the edge-to-edge spacing is 3.5 $\mu$m, and the coupling-zone length is 0.2 mm. The coupling loss for the 780-nm signal is as low as ~1\%, while the pump coupling efficiency reaches ~90\%. After the directional coupler, the pump and signal fields are combined and launched into the quasi-phase-matched (QPM) waveguide, which has a width of 8 $\mu$m and a poling period of 16.52 $\mu$m. At the chip output, a mode filter with a width of 6 $\mu$m is used to optimize the free space to fiber coupling efficiency. An two-channel fiber array containing an HI780 single mode fiber and an SMF-28 Ultra single mode fiber was pigtailed to the waveguide input port, and the overall transmission efficiency is approximately 65\% for the 780-nm signal and 50\% for the 1917-nm pump.

The down-converted 1315-nm photons are collected using an AR-coated aspheric lens. A dichroic mirror (reflectivity $>$99.8\% at 1315 nm and transmission $>$92\% at 1917 nm) is used to separate the residual pump and the 1315-nm photons. The 1315-nm photons then pass through a cascade of spectral filters: a long-pass filter (cut-on 1150 nm), a short-pass filter (cut-off 1370 nm), and a narrow-band filter (1-nm FWHM), with a combined transmission of ~97\%. These filters suppress parasitic nonlinear noises associated with difference-frequency generation (DFG), such as spontaneous Raman scattering and second- and third-harmonic generation.

\begin{figure}[htbp]
	\centering
	\includegraphics[width=.9\linewidth]{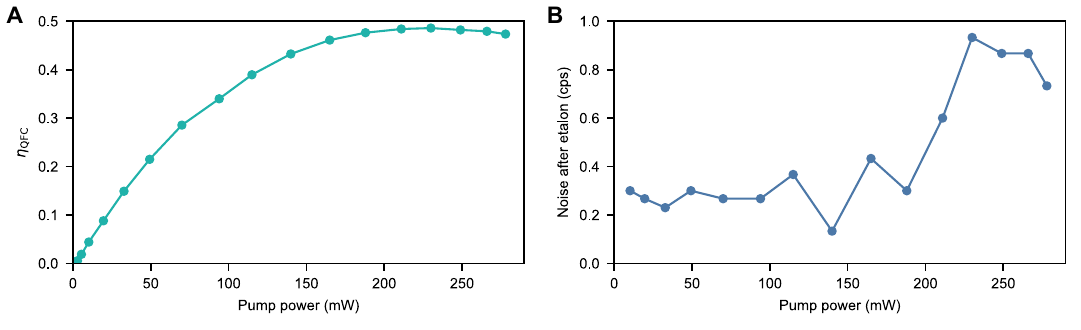}
	\caption{\textbf{Quantum frequency conversion performance.} 
		\textbf{A.} Conversion efficiency of the QFC Module as a function of pump power. The transmission of filter cavity is not included.
		\textbf{B.} Measured noise counts with 10 m fiber after cavity filtering.}
	\label{fig:QFC2}
\end{figure}

The measured overall efficiency of QFC module--including conversion, filtering, and fiber coupling--is 47\% at a pump power of 230 mW. To further suppress nonlinear noises introduced during long-fiber transmission, additional filter cavities (FWHM $\approx$ 300 MHz, transmission $\approx$ 85\%) and narrow-band filters (1-nm bandwidth) are inserted after Charlie’s beam splitter. The measured efficiency and noise count rate with pump power are shown in Fig. ~\ref{fig:QFC2}.

\subsection{Hyperfine Raman Control}

Our Raman laser system follows the scheme of Ref.~\cite{levine2022dispersive}, which employs a passive dispersive element to convert phase modulation into amplitude modulation. In our implementation, a 795-nm diode laser (TOPTICA TA-PRO) is phase-modulated by a resonant electro-optic modulator (HYQ Co., Ltd., HYQ-EOM-6.835G) driven at the hyperfine qubit frequency. The microwave drive for the EOM is generated by an arbitrary waveform generator (Zurich Instruments SHFSG+), providing programmable control over both the amplitude and phase of the modulation. For real-time basis selection during DI-QKD measurements, the microwave phase and laser amplitude are set on a shot-by-shot basis within 200~ns by a home-built quantum random number generator (QRNG).

The Raman beams are detuned by 80~GHz to the blue of the D$_1$ transition with circular polarization (Fig.\ref{fig:Raman}.A). Pulses which drives $\Delta m_{F} = \pm1$ transitions are applied perpendicular to the bias magnetic field, while pulses which drives $\Delta m_{F} = 0$ transitions are incident along the quantization axis. Because the qubit subspace is magnetically sensitive, a residual d.c. component of the perpendicular Raman fields can induce differential light shifts, effectively rotating the quantization axis defined by the bias magnetic field and causing leakage outside the qubit subspace. To suppress this effect, we shape the Raman pulse power with a Gaussian temporal envelope (FWHM 600 ns), which adiabatically reduces off-resonant coupling. Leakage errors are kept below 0.01\% per $\pi$-pulse, yielding qubit rotations with fidelities exceeding 99.9\% (Fig.~\ref{fig:Raman}B).

\subsection{Qubit coherence}
In our experiment, the qubit is encoded in the magnetically sensitive states $\ket{F = 2, m_F = -2}$ and $\ket{F = 1, m_F = -1}$. The main sources of decoherence arise from magnetic-field fluctuations and spin-motion coupling in the tightly focused optical tweezers~\cite{thompson2013coherence}. To extend the qubit coherence time, we (1) actively stabilize the bias magnetic field to the milligauss level using feedback control, and (2) lower the trap depth to approximately 100 $\mu$K during entanglement generation, which reduces differential light shifts and motional dephasing. 
\begin{figure}[htbp]
	\centering
	\includegraphics[width=.75\linewidth]{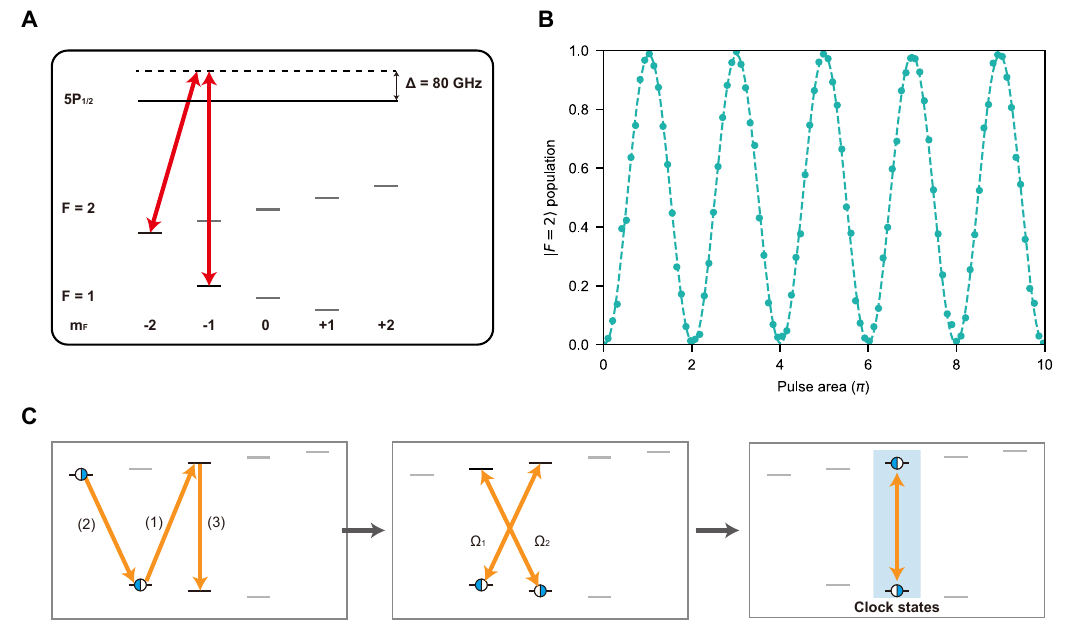}
	\caption{\textbf{Ground state Raman control.}  
		\textbf{A.} The ground-state hyperfine levels $F=1$ and $F=2$ of the $5S_{1/2}$ manifold are coupled via a virtual level detuned by $\Delta = 80$~GHz from the $5P_{1/2}$ excited state. Red arrows indicate the Raman transition pathways between Zeeman sublevels $m_F$.
		\textbf{B.} Rabi flopping between $\ket{\uparrow}$ and $\ket{\downarrow}$.
		\textbf{C.} Sequences to transfer qubit to clock states.
	}
	\label{fig:Raman}
\end{figure}

To further extend coherence time, we employ dynamical decoupling (DD) sequences that refocus low-frequency noise and suppress residual dephasing. Since the qubit resonance frequency shifts depending on whether the Raman beams are present, due to differential light shifts, the phases of the Raman pulses are carefully calibrated to compensate for these shifts, ensuring that each $\pi$-pulse remains resonant and that the self-correcting XY-4 and XY-8 sequences operate effectively. 

The qubit coherence time is further extended by transferring the population into magnetically insensitive clock states (Fig.\ref{fig:Raman}.C), where the first-order Zeeman shift vanishes. The pulse sequences are shown in Fig. S2. First, $\ket{\uparrow}$ and $\ket{\downarrow}$ are transferred to $\ket{F = 1, m_F = -1}$ and $\ket{F = 1, m_F = 0}$, respectively. Subsequently, a Raman pulse simultaneously couples the transitions $\ket{F = 1, m_F = -1} \leftrightarrow \ket{F = 2, m_F = 0}$ and $\ket{F = 2, m_F = -1} \leftrightarrow \ket{F = 1, m_F = 0}$ with Rabi frequencies satisfying $\Omega_1 / \Omega_2 = \sqrt{3}$. By setting the pulse duration $T$ such that $\Omega_2 T = 4$ and $\Omega_1 T = 4\sqrt{3} \approx 6.93$, the population is coherently mapped onto the clock-state pair $\ket{F = 1, m_F = 0}$ and $\ket{F = 2, m_F = 0}$. The transfer fidelity reaches approximately 0.97, limited primarily by the non-integer pulse area ratio and by differential a.c. Stark shifts induced by $\Omega_1$ and $\Omega_2$. The coherence time of clock state is measured 300 ms with XY-8-32 sequences, which can be extended over 1s with more $\pi$ pulses.

\subsection{Rydberg control}
The Rydberg state is highly sensitive to stray electric fields, which can induce energy level shifts and degrade coherence. To mitigate these effects, the atoms are confined within a titanium vacuum chamber equipped with indium-tin-oxide (ITO)-coated optical viewports. In addition, the internal surface of the in-vacuum aspherical lens are ITO-coated. Together, these conductive layers effectively form a Faraday cage around the atoms, suppressing charge accumulation on dielectric surfaces and shielding the system from external field fluctuations. Using Rydberg Stark spectroscopy, we estimate the residual stray electric field at the atom position to be below 15 mV/cm, ensuring stable and reproducible Rydberg excitation over hundreds of experimental hours. The small residual frequency difference between the two memory nodes is compensated by in-vacuum octupole electrodes, which allow fine-tuned electric-field balancing and long-term frequency alignment.

\begin{figure}[htbp]
	\centering
	\includegraphics[width=.9\linewidth]{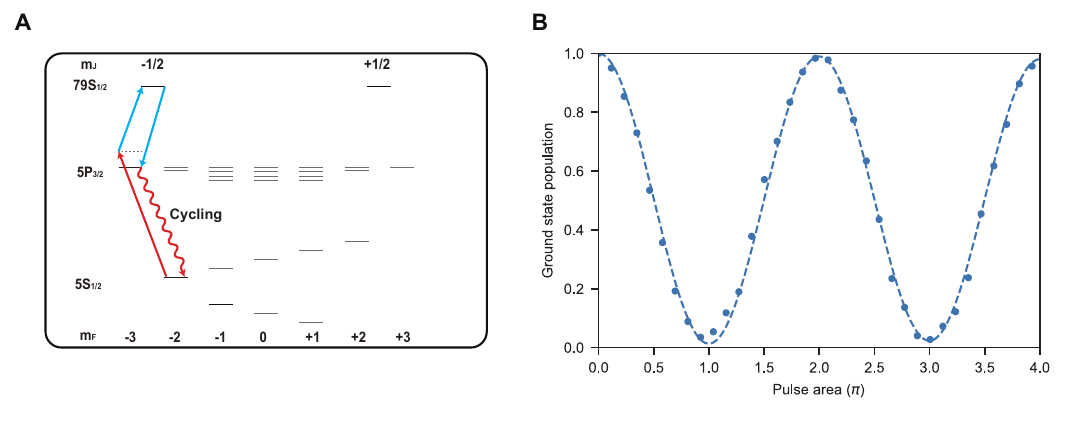}
	\caption{\textbf{Rydberg control.}  
		\textbf{A.} Energy diagram of Rydberg excitation and de-excitation.
		\textbf{B.} Rabi flopping between $\ket{\uparrow}$ and $\ket{r}$.
	}
	\label{fig:Rydberg}
\end{figure}

The ground-to-Rydberg transition in ${}^{87}\mathrm{Rb}$ is driven via a two-photon excitation scheme using a 780-nm laser (Precilaser FL-SF-780-0.2-CW) and a 479-nm laser (Precilaser FL-SF-480-2.5-CW) (Fig.~\ref{fig:Rydberg}.A). The 479-nm light is generated by cascaded frequency doubling of a 1917-nm fiber laser, which also serves as the pump for the QFC. Both excitation beams are frequency-stabilized by locking to a high-finesse reference cavity, resulting in linewidths below 100 kHz over experimental timescales. At the atom position, the beams are arranged in a counter-propagating geometry along the quantization axis to suppress Doppler broadening, with beam waists of $100~\mu$m (780 nm) and $15~\mu$m (479 nm). The corresponding Rabi frequencies are $\Omega_{780} = 2\pi\times50$ MHz and $\Omega_{479} = 2\pi\times55$ MHz. The excitation is detuned by 400 MHz to the blue of the D$_2$ line, yielding a two-photon Rabi frequency of approximately $2\pi\times3.5$ MHz.

Population transfer from the Rydberg state $\ket{r}$ to the excited state $\ket{e}$ is achieved by applying an additional resonant 479-nm pulse delayed by 150 ns after the initial excitation. The transfer pulses, generated by an AOM and Pockels cell, have a Gaussian temporal profile with a 10-ns full width at half maximum, enabling fast and coherent mapping. The transfer fidelity is measured to be approximately 99.5\%, limited primarily by the finite Rydberg-state lifetime and residual laser-power fluctuations.

The coherence between the ground state and the Rydberg state is characterized using Ramsey interferometry. Two $\pi/2$ pulses separated by a variable delay are applied, and the resulting interference fringes reveal a coherence time of 12~$\mu$s, limited primarily by the finite atomic temperature. This coherence time is sufficient for the excitation and transfer sequence and therefore introduces no significant decoherence in the entanglement process.
 
\subsection{Phase stabilization}
We first review the phases contributing to the atom-atom entangled state. The atom-photon entangled state prior to the optical BSM can be written as
\begin{equation}	
	\psi_{ap} = \sqrt{1-\alpha}\ket{\downarrow}\ket{0}+\text{e}^{i\phi}\sqrt{\alpha}\ket{\uparrow}\ket{1}
\end{equation}
where the phase factor $\phi$ arises from several physical processes.

\textbf{Raman rotations and state evolution $\phi_{atom}$}. During atom-photon entanglement (APE) generation, multiple Raman pulses are applied to manipulate the atomic qubit. These operations imprint the phase of the microwave drive field onto the atomic state, thereby contributing directly to $\phi$. In addition, the subsequent free evolution of the qubit—including both ground-state evolution and ground-Rydberg superpositions—accumulates a deterministic phase due to Larmor precession. Together, these contributions form a well-defined phase denoted as $\phi_{\text{atom}}$.

\begin{figure}[htbp]
	\centering
	\includegraphics[width=.9\linewidth]{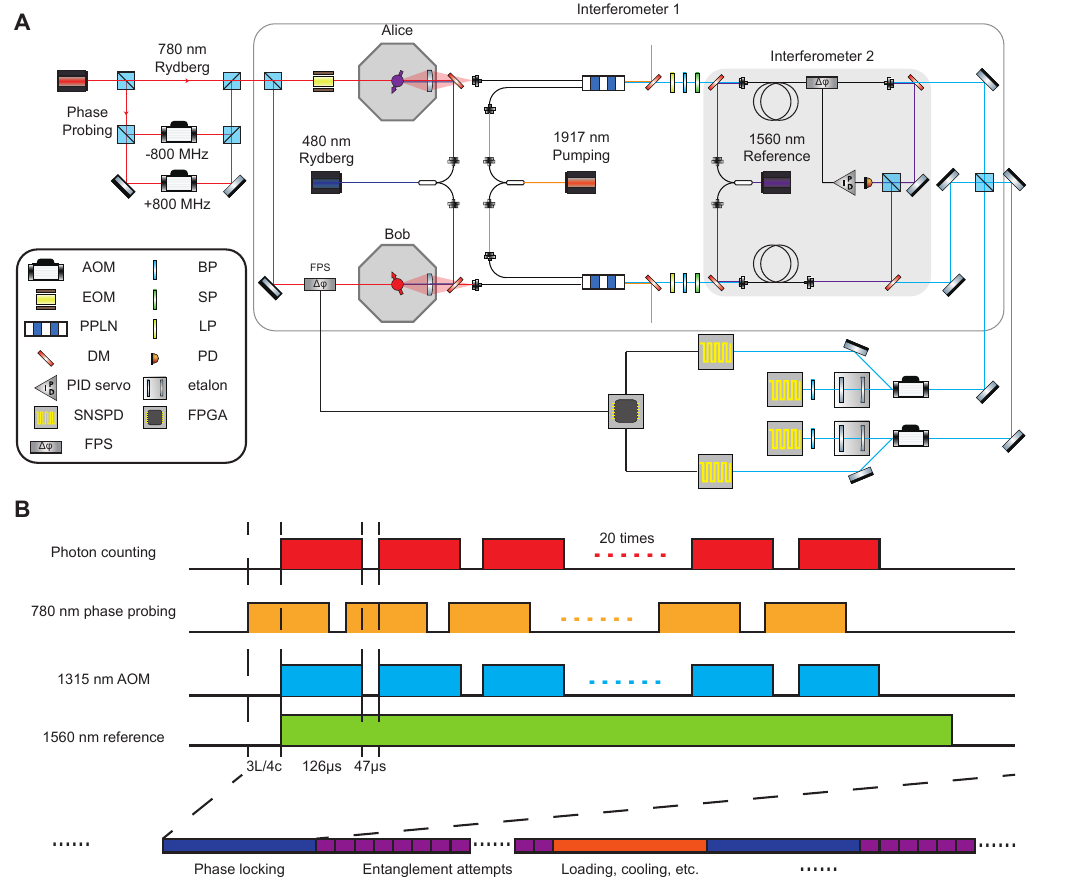}
	\caption{\textbf{Phase stabilization.} 
		\textbf{A.} Schematic of the dual-interferometer system with Rydberg excitation lasers, QFC modules, telecom links, and feedback loops for active phase stabilization.
		\textbf{B.} Sequences for phase stablization. Phase probing is pulsed before entanglement attempts, while 1560 nm reference beams are holding on during the entanglement attempts.}
	\label{fig:phase_setup}
\end{figure}

\textbf{Rydberg excitation and de-excitation $\phi_{\text{Ryd}}$.} The optical phases of the 780-nm and 479-nm Rydberg excitation and de-excitation pulses are coherently imprinted onto the spin-photon entangled state. Consequently, their relative phase contributes directly to the overall phase $\phi$, which we denote as $\phi_{\text{Ryd}}$:
\begin{equation}
	\phi_{\text{Ryd}} = k_{780}L_{780} + k_{479,e}L_{479} - k_{479,d}L_{479} + \bigl(\phi_{780} + \phi_{479,e} - \phi_{479,d}\bigr)
\end{equation}
where $k_{780}$ is the wave vector of the 780-nm Rydberg laser, and $k_{479,e}$ ($k_{479,d}$) are the wave vectors of the 479-nm excitation (de-excitation) lasers. $L_{780}$ and $L_{479}$ denote the corresponding optical path lengths from the lasers to the atom, while $\phi_{780}$, $\phi_{479,e}$, and $\phi_{479,d}$ represent their respective phase offsets. Here, the dependence on the 479-nm beams is expressed through the effective wave vector associated with the intermediate detuning, defined as $k_{\Delta} = k_{479,e} - k_{479,d}$.

\textbf{Optical path of the signal photon $\phi_{\text{sig}}$.}
The signal photon generated by the atom is first coupled into a short Hi780 fiber and then directed into the QFC module. After frequency conversion, it propagates through the long telecom fiber $L_{\text{tel}}$ before interfering with the photon from the other node at the BSM station. The accumulated relative phase is given by
\begin{equation}
	\phi_{\text{sig}} = k_{\text{pho}} L_{\text{signal}} + k_{\text{tel}}(L_{\text{wg}} + L_{\text{tel}})
\end{equation}
where $k_{\text{pho}}$ and $k_{\text{tel}}$ are the wave vectors of the emitted 780-nm photon and the down-converted 1315-nm photon, respectively. Here $L_{\text{signal}}$ is the optical length from atom to  waveguide's input port, $L_{\text{wg}}$ is the effective optical length of the waveguide, and $L_{\text{tel}}$ is the optical length length from the waveguide output to the BSM, including the long telecom fiber.

\textbf{QFC pumping $\phi_{\text{QFC}}$.} In QFC process, the optical phase of pumping laser will also be imprinted into $\phi$:
\begin{equation}
	\phi_{\text{pump}} = k_{\text{pump}}L_{\text{pump}} + \phi_{pump}
\end{equation}
where $k_{\text{pump}}$ is the wave vector of the pump laser and $L_{\text{pump}}$ is the corresponding optical path length. The constant offset term $\phi_{pump}$ denotes respective phase offset.

The totol phase of the APE state is thus $\phi = \phi_{\text{atom}} + \phi_{\text{Ryd}} + \phi_{\text{sig}} + \phi_{\text{QFC}}$. After a successful BSM, the generated atom-atom entanglement reads:
\begin{equation}
	\psi_{AB} = \frac{1}{\sqrt{2}}(\ket{\downarrow\uparrow}\pm\text{e}^{i(\phi_{A}-\phi_{B})}\ket{\uparrow\downarrow})
\end{equation}
From this expression it is clear that only the relative phase difference $\phi_{A}-\phi_{B}$ between the two nodes affects the entangled state. Since the 780-nm and 479-nm Rydberg lasers, as well as the QFC pump laser, are shared between both nodes, their phase offsets $\phi_{780}$, $\phi_{479,e}$, $\phi_{479,r}$ and $\phi_{pump}$ are identical and therefore cancel in the relative phase. These contributions can thus be neglected in the following analysis. Furthermore, the phase $\phi_{atom}$ arising from Raman operations and atomic evolution is deterministic and well defined, and can likewise be ignored in the relative-phase considerations. The phase difference $\delta\phi = \phi_{A}-\phi_{B}$ is thus:
\begin{equation}
	\begin{aligned}
		\delta\phi &= k_{780}(L_{780,A}-L_{780,B}) + k_{\Delta}(L_{479,A}-L_{479,B}) + k_{pho}(L_{sig,A}-L_{sig,B})\\
		&+k_{tel}(L_{wg,A}-L_{wg,B}+L_{tel,A}-L_{tel,B})+k_{pump}(L_{pump,A}-L_{pump,B})
	\end{aligned}
\end{equation}
Since the optical path length for the 479-nm beams is only about 10 m, the phase variation from $(L_{479,A}-L_{479,B})$ is negligible compared with the effective wavelength of 0.75 m, and can therefore be safely ignored. Similarly, the frequency difference between the 780-nm photon and 780-nm Rydberg laser is $\Delta$,so this contribution can be also ignored:
\begin{equation}
	\begin{aligned}
		\delta\phi &= k_{pho}(L_{780,A}-L_{780,B}+L_{sig,A}-L_{sig,B})\\&+k_{tel}(L_{wg,A}-L_{wg,B}+L_{tel,A}-L_{tel,B})+k_{pump}(L_{pump,A}-L_{pump,B})
	\end{aligned}
\end{equation}
Thus, active stabilization is required only for the 780-nm Rydberg laser and telecom channels, corresponding to stabilizing the phase of a Mach-Zehnder interferometer, as Shown in Fig.\ref{fig:phase_setup}. Prior to the entanglement generation, weak phase-probing pulses (at a count rate of 2 MHz) through the same optical paths as the 780-nm Rydberg lasers. To avoid heating during phase locking, as well as lock-point drift caused by fiber-length fluctuations, the phase-probing pulses are detuned by $\pm$ 800 MHz from atomic resonance~\cite{luo2025entangling}. These pulses are subsequently coupled into the fiber leading to the QFC module, with a coupling efficiency of approximately 0.6\%. An FPGA processes the photon counts from SNSPD C/D to calculate the phase difference, and the resulting error signal is fed back to the fiber stretcher for active phase stabilization. This feedback operation is repeated 20 times to ensure robust convergence of the phase lock before entanglement attempt.

\begin{figure}[htbp]
	\centering
	\includegraphics[width=.9\linewidth]{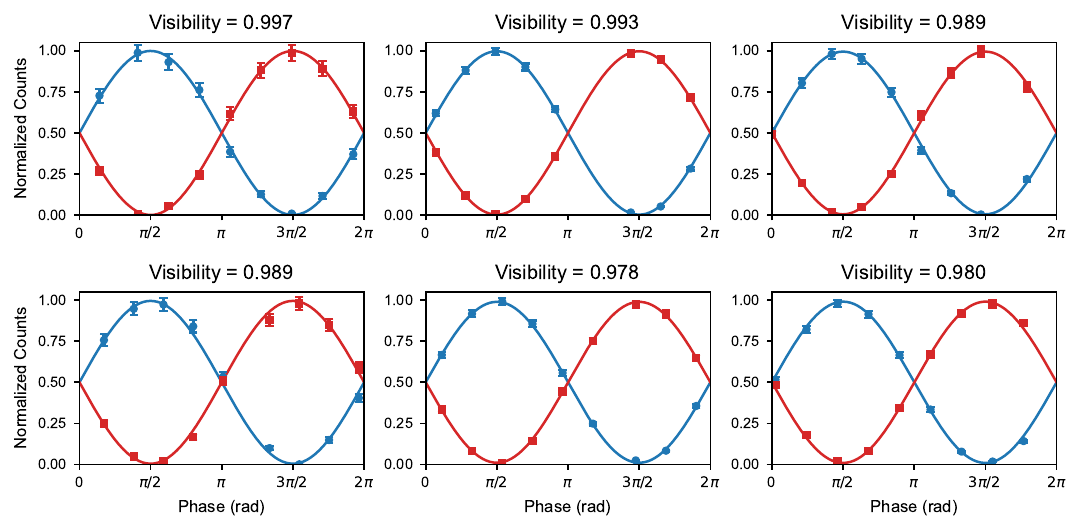}
	\caption{\textbf{Measured interference of weak coherent pulses.} 
		Interference fringes recorded after transmission through different fiber lengths: 
		(Top left) 0 km (without QFC, visibility = 0.997), 
		(Top middle) 11 km (0.993), 
		(Top right) 20 km (0.989), 
		(Bottom left) 50 km (0.989), 
		(Bottom middle) 70 km (0.978), 
		(Bottom right) 100 km (0.980). 
		The visibilities remain above 0.98, confirming high phase stability of the interferometer over long distances.}
	\label{fig:phase_result}
\end{figure}

During the entanglement attempts, the active weak-probe feedback loop is disabled to avoid introducing additional noise or spurious counts. Nevertheless, the relative phase of the fibers drifts rapidly, making passive stabilization insufficient. To mitigate this, another 1560-nm locking loop is activated 4 ms before each experimental cycle and deactivated upon completion. The resulting signal is monitored by a photon detector. A PID servo processes this error signal and drives another fiber stretcher to compensate fast phase drifts,thereby maintaining telecom-fiber phase stability throughout the entanglement sequence. 

To probe phase stability, weak coherent pulses resonant with the single-photon frequency are sent through the interferometer as analogs of the generated single photons. The repetition rate and total number of weak pulses are matched to those of the entanglement generation attempts. By scanning the locking point of the Mach-Zehnder interferometer, the resulting interference fringes are recorded, and the fringe visibility is extracted as a quantitative measure of phase stability, as shown in \ref*{fig:phase_result}. The measured visibilities remain above 0.985 over fiber lengths up to 100 km, confirming that the implemented stabilization scheme effectively suppresses phase noise over the entire experimental distance.

\subsection{Photon recoil and Young's double slit analogy}
The main source of error in atom-atom entanglement generated via the SPI scheme arises from residual photon recoil during the excitation and emission processes. When a photon is emitted, the atomic qubit acquires a momentum kick that entangles its internal spin state with motional degrees of freedom, thereby reducing the fidelity of the spin-photon entangled state. During the RSPE process, the optical tweezer is switched off to avoid light shifts on the Rydberg states. The atomic wave packet expands during the release time, and the coupling between atomic motion and photon recoil for an expanded wave packet can be described by the Debye-Waller factor $D(t)$:
\begin{equation}
	D(t) = \prod_{i = x,y,z}\exp{(-\eta_{i}^2(1+\omega_i^2 t^2))}
\end{equation}
where $t$ denotes the release time from the optical tweezer, and $\eta_i = Q_i r_i(t=0)$ is the Lamb-Dicke parameter, with $Q_i = \vec{k}_{i,1} + \vec{k}_{i,2}-\vec{k}_{i,3}-\vec{k}_{i,4}$ representing the net momentum transfer during the RSPE process along axis $i$. Here, $r_i(t=0)$ is the initial ground-state wave-packet size along axis $i$, and $\omega_i$ is the trap frequency before release. Since the trap frequency is approximately 30 kHz, the characteristic motional period of the atom is around 30~$\mu$s, which is many orders of magnitude longer than the duration of the RSPE process (300 ns). Therefore, the time dependence of the wave-packet expansion can be neglected, and $D(t)$ can be approximated using the initial confinement.

In the Young's double-slit analogy presented in the main text, the qubit is initially prepared in $\ket{\uparrow}$. The total state of atom and photon before the RSPE process can be written as
\begin{equation}
	\psi_0 = \ket{\uparrow}\ket{\xi_0}\ket{0}_{p},
\end{equation}
where $\ket{\xi_0}$ denotes the initial atomic momentum state. By controlling the pulse area of the Rydberg transition, we excite a fraction $p = 0.02$ of the population to $\ket{r}$, and subsequently transfer it to $\ket{e}$. After the RSPE process, the joint state becomes
\begin{equation}
	\psi =\ket{\uparrow}\otimes[\ket{\xi_0}(\sqrt{1-p}\ket{0}_{p}+\exp{(i\varphi)}\sqrt{pD}\ket{1}_{p})+\exp{(i\varphi)}\ket{\xi'}\sqrt{p(1-D)}\ket{1}_{p}]
\end{equation}
where $\ket{\xi'}$ denotes orthogonal states of $\ket{\xi_0}$, and $\varphi$ is the phase introduced by the Rydberg lasers. The Debye-Waller factor $D$ describes the probability that, during the RSPE process, the atomic motional state remains unchanged and no transition between different momentum states occurs, analogous to the zero-phonon line observed in solid-state defects, where optical transitions occur without creating or absorbing lattice phonons. Correspondingly, $1-D$ represents the probability that photon recoil induces a phonon jump. In this case, the atomic momentum carries which-way information, making it, in principle, possible to determine from which node the single photon comes. This residual 'distinguishability' reduces the interference visibility at the beam splitter and therefore lowers the fidelity of the heralded entangled state.

At Charlie's beam splitter, photons from Alice and Bob interfere with each other. The density matrix of photons is obtained by tracing over atomic spin and motional states, which gives:
\begin{equation}
	\rho_p = 
	\begin{pmatrix}
		p & \exp{(i\phi)}\sqrt{p(1-p)D} \\
		\exp{(-i\phi)}\sqrt{p(1-p)D} & 1-p
	\end{pmatrix}
\end{equation}
which leads to a maximal single-photon interference visibility of $V \approx D(1-p)$. For simplicity, the Rydberg excitation lasers can be regarded as plane waves, given their large beam waists. Considering the specific mode defined by the aspherical lens collection and fiber coupling, as well as the effect of finite atomic temperature, the resulting Debye-Waller factor is estimated to be $D \approx 0.955$. Together with $p = 0.02$, this yields a predicted visibility of $V \approx 0.935$, which is in good agreement with our experimentally observed single-photon interference contrast. For SPI-based optical BSM, this predicted visibility directly determines the BSM fidelity. The Debye-Waller factor $D$ can be further increased by lowering the atomic temperature to reduce atomic wave-packet spread, enhancing trap confinement to enter deeper into the Lamb-Dicke regime, and lowering the numerical aperture of the collection optics.

\subsection{Fidelity of atom-atom entanglement}
The main error sources affecting atom-atom entanglement fidelity arise from double excitation, residual photon recoil, state preparation and measurement (SPAM) errors, finite atomic cohere-nce, phase drift of the interferometer, and background noise. The error budget are listed in Tab~\ref{tab:sup_error_budget}.

\begin{table}[htbp]
	\centering
	\caption{\textbf{Error budget for atom–atom entanglement fidelity at different fiber lengths.}
	}
	\label{tab:sup_error_budget}
	
	\begin{tabular}{|c|c|c|c|c|c|}
		\hline
		\textbf{Source of errors} & \textbf{11 km} & \textbf{20 km} & \textbf{50 km} & \textbf{70 km} & \textbf{100 km} \\
		\hline
		double excitation &0.023&0.028&0.033&0.035&0.038\\
		\hline
		photon recoil &0.022&0.022&0.022&0.022&0.022\\
		\hline
		SPAM and single-qubit rorations &<0.002&<0.002&<0.002&<0.002&<0.002\\
		\hline
		atomic coherence &<0.001&<0.001&<0.001&<0.001&<0.002\\
		\hline
		phase drift of the interferometer &<0.003&<0.005&<0.005&<0.01&<0.01\\
		\hline
		background noise &<0.002&<0.002&0.005&0.0103&0.0285\\
		\hline
		\textbf{total} &<0.0530 & <0.0600 & <0.0680 & <0.0803 & <0.1025\\
		\hline
	\end{tabular}
\end{table}

\subsection{Experimental Time sequence}
\begin{figure}[htbp]
	\centering
	\includegraphics[width=.9\linewidth]{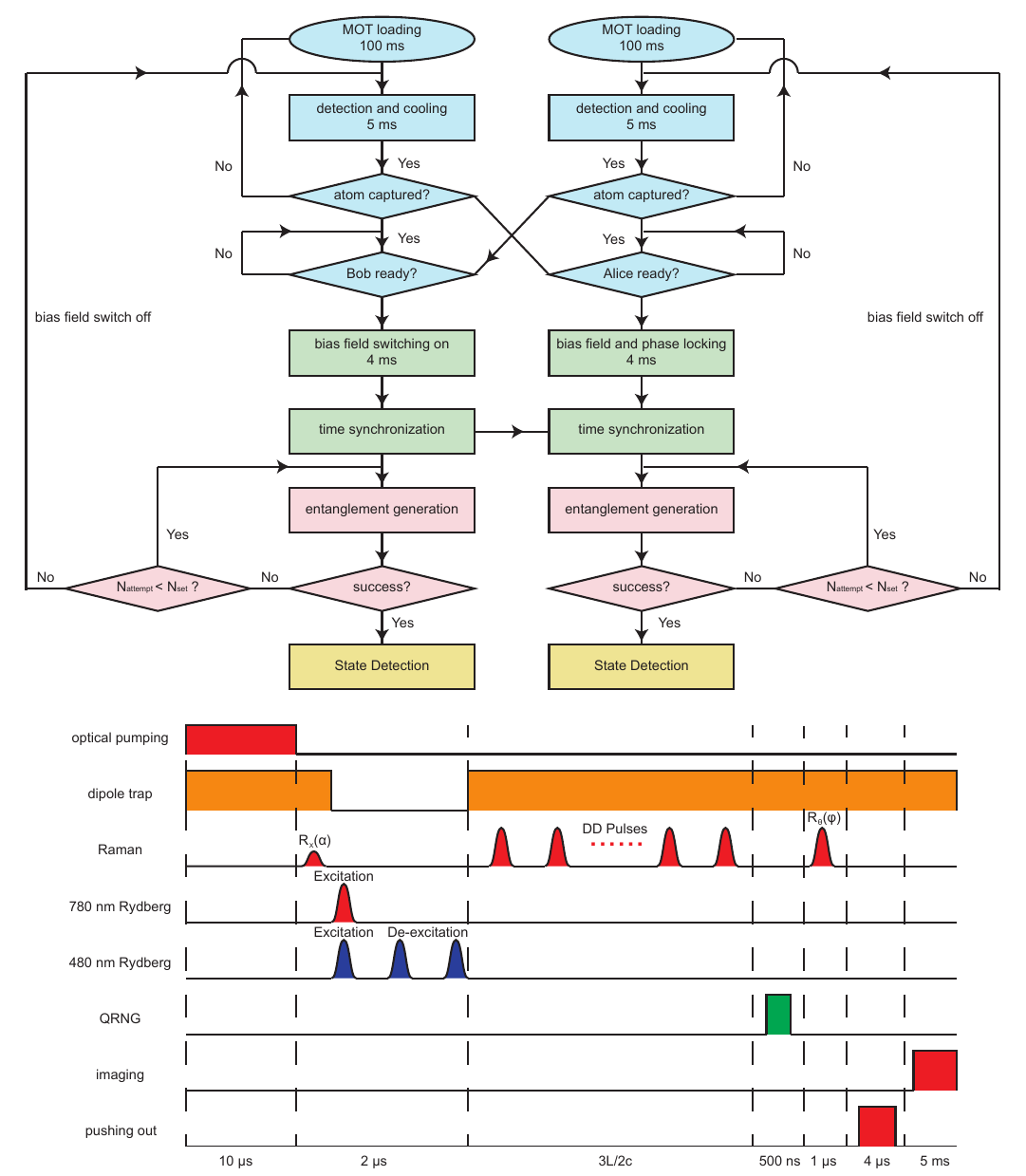}
	\caption{\textbf{Experimental sequence of atom–atom entanglement generation.}  
		\textbf{A.} Flowchart of MOT loading, atom preparation, bias-field switching, synchronization, entanglement attempts, and state detection.  
		\textbf{B.} Timing diagram of entanglement generation and state measurement.}
	\label{fig:time_sequence}
\end{figure}

The experimental sequence of atom-atom entanglement generation, as well as DI-QKD, is illustrated in \ref{fig:time_sequence}. It begins with magneto-optical trap (MOT) loading, during which the presence of a single atom in each optical tweezer is checked every 100 ms. Once an atom is successfully captured at one node, the system verifies the readiness of the partner node. When both nodes are prepared, they continue to prepare for entanglement generation phase begins. Within 4 ms, the bias magnetic field is ramped up and stabilized at 6.6 G. During this interval, the polarization-locking module and the phase-locking systems for the 1560-nm and 1315-nm lasers are engaged and stabilized. At the same time, the 852-nm dipole-trap power is adiabatically ramped down to 100 $\mu$K over 700 $\mu$s, which suppresses parametric heating that would otherwise result from repeated trap switching during the subsequent Rydberg excitation stage.

Following preparation, the entanglement generation attempt begins. After a 10-$\mu$s optical pumping stage, a sequence of Rydberg excitation and de-excitation pulses is applied to generate single photons via the RSPE process. To clear any residual population in the Rydberg state, the de-excitation pulse is applied twice, with a peak-to-peak interval of 200 ns between successive pulses. After a delay of $3L/2c$, corresponding to the one-way propagation time of the photons through the fiber plus the classical communication time, the system checks for a heralding signal. If no heralding signal is observed, the system resets and initiates another entanglement attempt, repeating the cycle until either a successful heralding event occurs or a predefined timeout is reached. This timeout is set to 2.8 ms, which is determined by the phase-stability window of the interferometer. Once the timeout is reached, both nodes return to atom-presence checks and cooling before the next cycle begins. If a heralding event is registered, the memory qubits are projected into an entangled state and the protocol proceeds directly to measurement. At this point, the QRNG is triggered in real time, generating a two-bit random number within 300 ns to determine the measurement basis. A corresponding Raman pulse is then applied to implement the chosen basis rotation, after which state readout is performed via fluorescence detection. The outcome is recorded and stored for subsequent key-rate analysis in the DI-QKD protocol. 

\subsection{Entangling rate and feasible improvement}
\subsubsection{Entangling rate}
The entanglement generation rate $R_{\text{ent}}$ in the SPI scheme is given by $R_{\text{ent}} = p_s R_{\text{rep}}$,  
where $p_s = \alpha_A \eta_A + \alpha_B \eta_B$ denotes the success probability per trial. Here, $\eta_{A(B)}$ represents the total transmission efficiency from node A (B) to the Bell-state measurement (BSM) station—including coupling, frequency conversion, fiber transmission, and detector efficiencies—while $R_{\text{rep}}$ is the experimental repetition rate. In the experiment, both $\alpha$ and $\eta$ are well balanced between the two nodes.

The duty cycle of entanglement generation attempts is about 0.15, with a time overhead per trial of 12 $\mu$s + 3L/2c. The duty cycle is limited by several issues:

\textbf{Communication time.} Photon propagation and classical heralding signals introduce an unavoidable latency, setting the fundamental upper bound on the repetition rate. This limitation can, in principle, be overcome through temporal multiplexing.

\textbf{Magnetic-field switching}. In the experiment, polarization-gradient cooling requires a near-zero magnetic field, so the bias field must be switched off during cooling and then ramped up and stabilized before each group of entanglement attempts. Eddy currents in nearby conductive components, particularly the vacuum chamber, slow the field response and impose an overhead of about 5 ms per cycle. This limitation could be alleviated by relocating the coil inside the vacuum to suppress eddy currents, or by implementing cooling directly at a finite bias field, thereby avoiding repeated switching.

\textbf{Atom loading.} The average loading time of a single atom into the optical tweezer is on the order of hundreds of milliseconds, which dominates the experimental duty cycle. Incorporating an atomic reservoir would significantly reduce the loading time. 

\textbf{State detection.} The push-out method provides high-fidelity state discrimination but is destructive, necessitating atom reloading after each measurement. This limitation could be overcome in the future by implementing nondestructive state readout.

\begin{table}[htbp]
	\centering
	\caption{Optical efficiencies in the experiment. Fibre transmission is listed for different link lengths.}
	\begin{tabular}{l c}
		\hline
		Component & Efficiency (\%) \\
		\hline
		Collection by aspheric lens & 8.5 \\
		Fluorescence fiber coupling & 50 \\
		QFC module                 & 47 \\
		Insertion loss of polarization controller and fiber stretcher   &86   \\
		BSM Node (including filter cavity)            	& 76.5 \\
		SNSPD detection efficiency & 85 \\
		\hline
		Fibre transmission (11 km /2) & 68.3 \\
		Fibre transmission (20 km /2) & 46.7 \\
		Fibre transmission (50 km /2) & 15.0 \\
		Fibre transmission (70 km /2) & 6.9 \\
		Fibre transmission (100 km /2) & 2.2 \\
		\hline
		Total efficiency (11 km)   & 0.72 \\
		Total efficiency (20 km)   & 0.49 \\
		Total efficiency (50 km)   & 0.16 \\
		Total efficiency (70 km)   & 0.072 \\
		Total efficiency (100 km)  & 0.023 \\
		\hline
	\end{tabular}
\end{table}

\subsubsection{Feasible improvement}
For realistic DI-QKD implementation, the entangling rate demonstrated in this work is relatively low. Several feasibility improvements can help increase the entanglement generation rate:

\textbf{Continuous Operation.} The repetition rate of entanglement generation is currently limited by several technical factors, including pulsed reloading of atoms, magnetic field switching, and destructive state readout. Recently, a continuous operation architecture has been demonstrated in large-scale neutral atom-based quantum computation, where atom loading, qubit operations, and state readout are implemented in parallel. With this architecture, atoms no longer need to be reloaded, significantly increasing the repetition rate of entanglement generation.

\textbf{Multiplexing.} In the current experiments, each node is equipped with only a single atom, which severely limits the entanglement generation rate. Using atom array, once a single emitter has emitted a photon, the system can quickly switch to the next emitter to generate entanglement, significantly boosting the entanglement generation. When combined with the continuous operation architecture, the repetition rate of the experiment is essentially limited only by the time required for a single entanglement generation trial ($\sim$500 ns), leading to a repetition rate of over 1 MHz.

\textbf{Integration of optical cavities.} In the present apparatus, the atom is coupled to free-space optics, which sets a fundamental limit on the source efficiency. By integrating optical cavities, the coupling efficiency between the atoms and the emitted photons can be significantly enhanced, increasing the source efficiency to over 40\%.

\textbf{Reducing link loss.} By converting the photon frequency to the 1.5 $\mu$m telecom band and using hollow-core fibers, the link loss can be reduced to as low as 0.09 dB/km. This approach significantly lowers transmission losses, particularly for long-distance communication.

With these improvements, the entanglement rate over a 10 km distance can exceed 10 kHz, corresponding to a key rate of over 1 kbit/s. The secure key rate and entanglement distance can be further extended by employing quantum networks with parallel quantum channels and quantum repeaters.

\subsection{Quantum random number generator}
In the DI-QKD experiment, measurement bases are chosen using a home-built quantum random number generator (QRNG). The chip (Fig. S7) integrates a laser diode, beam splitter, photodiodes, ADC, and FPGA, and exploits vacuum-state fluctuations as the entropy source. A strong local oscillator interferes with vacuum at the beam splitter, and balanced homodyne detection measures the resulting quadrature noise. The noise is Gaussian, with quantum and classical components distinguished by their scaling with local oscillator power. From this, the min-entropy is estimated, and a Toeplitz-matrix extractor generates uniform random bits with an extraction ratio of 0.67, balancing security and efficiency. To verify randomness quality, the QRNG output is tested with the NIST SP 800-22 statistical suite, which it passes successfully, confirming suitability for DI-QKD applications.

\begin{figure}[htbp]
	\centering
	\includegraphics[width=.9\linewidth]{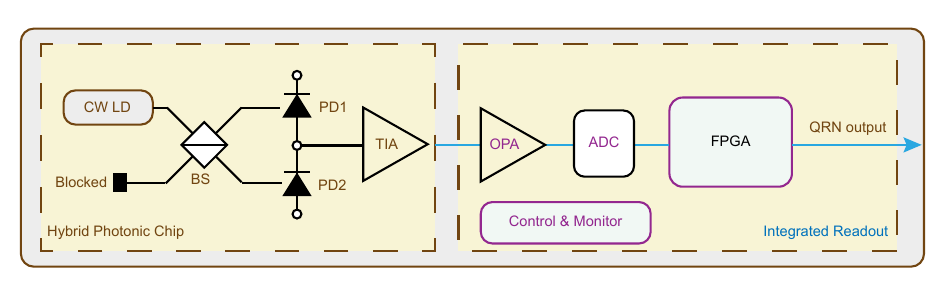}
	\caption{\textbf{Integrated quantum random number generator (QRNG) architecture.} A continuous-wave laser diode (CW LD) feeds a beam splitter (BS) on a hybrid photonic chip, with one input blocked. The two output ports are detected by photodiodes (PD1, PD2), and the difference current is amplified by a transimpedance amplifier (TIA). The signal is further amplified (OPA), digitized by an analog-to-digital converter (ADC), and processed on an FPGA to produce the QRN output. Integrated control and monitoring ensure stable operation.}
	\label{fig:QRNG}
\end{figure}

\subsection{Security proof and key rate of DI-QKD}
In the DI-QKD experiment, we use both the standard entropy accumulation theorem (EAT) and an improved version based on R\'{e}nyi entropy to estimate the secure key rate. In this section, we briefly outline the theoretical framework underpinning the security analysis, summarizing the key assumptions, the entropy bounds derived in each approach, and their implications for finite-size performance.

\subsubsection{The CHSH game}
The CHSH (Clauser-Horne-Shimony-Holt) game~\cite{CHSH} is a pivotal thought experiment in quantum mechanics, designed to investigate the non-local correlations predicted by quantum theory. In addition, it is shown that the game can be used in quantum communication protocols. The game involves two spatially separated players, Alice and Bob, each in possession of some system that could be correlated with the other. Alice and Bob independently and randomly select input bits $x,y\in\{0,1\}$ and perform a measurement on their system based on their input bit, producing output bits $a,b\in\{0,1\}$. The players win if their outputs satisfy $a\oplus b=x\land y$, where $\oplus$ denotes XOR and $\land$ denotes AND. 

Under local realism, the maximum probability of winning the CHSH game is bounded by 75\%, as dictated by the CHSH inequality, a mathematical constraint derived from the assumptions of locality (no faster-than-light signaling) and realism (preexisting measurement outcomes). Formally, the CHSH expression is 
\begin{equation}
S=E(0,0)+E(0,1)+E(1,0)-E(1,1),
\end{equation}
where $E(x,y)$ is the correlation between Alice’s and Bob’s outputs given inputs $x$ and $y$; local hidden variable theories enforce $|S|\leq2$. 

Quantum mechanics, however, violates this bound. By exploiting entanglement, Alice and Bob can choose measurement settings (e.g., along angles $0^{\circ},45^{\circ},90^{\circ},135^{\circ}$ in a Bloch sphere) to maximize $S$. For an entangled Bell state, the optimal quantum strategy yields $S=2\sqrt{2}\approx2.828$, exceeding the classical limit. Using this feature, the CHSH game serves as a benchmark for testing quantum nonlocality in experiments and underpins protocols like QKD, where security relies on the impossibility of local hidden variable theories to replicate quantum correlations. 

For the purpose of constructing a CHSH-based DI-QKD protocol, it is convenient to introduce the payoff function of the CHSH game as
\begin{equation}
\omega(a,b,x,y)=\begin{cases}
1, & \text{if }a\oplus b=x\land y,\\
0, & \text{otherwise}.
\end{cases}
\end{equation}

\subsubsection{The CHSH-based DI-QKD protocol}
We mostly follow the protocol proposed in~\cite{arnon2018practical}, though with slight modifications to include a sifting step to match the experimental setting, an explicit error verification step as discussed in~\cite{TSB+22}, and a simplification to the parameter estimation procedure suggested in~\cite{arq2024generalized}.

Given an expected CHSH winning probability $\omega_{\mathrm{exp}}$, two probabilities $\gamma_{A}$ and $\gamma_{B}$, and a small real number $\delta$ indicating whether the experimental data achieve $\omega_{\mathrm{exp}}$, the protocol goes as follows.
\begin{enumerate}
	\item Data Generation:
	For the $i\in\{1,\dots,n\}$ rounds of experiments:
	\begin{enumerate}
		\item Alice generates $s_{i}\in\{0,1\}$ according to probabilities $\{\gamma_{A},1-\gamma_{A}\}$. If $s_{i}=0$, Alice randomly and uniformly picks $x_{i}\in\{0,1\}$; if $s_{i}=1$, Alice picks $x_{i}=0$.
		\item Similarly, Bob generates $t_{i}\in\{0,1\}$ according to probabilities $\{\gamma_{B},1-\gamma_{B}\}$. If $t_{i}=0$, Bob randomly and uniformly picks $y_{i}\in\{0,1\}$; if $t_{i}=1$, Bob picks $y_{i}=2$.
		\item Alice and Bob measure their systems with measurement settings $x_{i}$ and $y_{i}$, and obtain measurement outcomes $a_{i}$ and $b_{i}$ with $a_{i},b_{i}\in\{0,1\}$, respectively.
		\item Alice and Bob have strings $\mathbf{X}=(x_{1}x_{2}\dots x_{n})$, $\mathbf{A}=(a_{1}a_{2}\dots a_{n})$, $\mathbf{S}=(s_{1}s_{2}\dots s_{n})$ and $\mathbf{Y}=(y_{1}y_{2}\dots y_{n})$, $\mathbf{B}=(b_{1}b_{2}\dots b_{n})$,  $\mathbf{T}=(t_{1}t_{2}\dots t_{n})$, respectively.
	\end{enumerate}
	\item Information reconciliation:
	\begin{enumerate}
		\item Alice and Bob exchange $\mathbf{X}$, $\mathbf{S}$ and $\mathbf{Y}$, $\mathbf{T}$ to each other, respectively.
		\item Alice and Bob set their bits $a_{i}=b_{i}=0$ if $(s_{i}, t_{i})=(1,0)$ and $(s_{i},t_{i},x_{i},y_{i})=(0,1,1,2)$.
		\item Bob further set his bits $b_{i}=0$ if $y_{i}=2$.
		\item Using an error-correction protocol, Bob encodes $\mathbf{A}$ to $\mathbf{L}=\mathrm{encoding}(\mathbf{A})$ and sends $\mathbf{L}$ to Bob, who decodes $\mathbf{L}$ and obtains a guess of $\mathbf{A}$ as $\hat{\mathbf{A}}$.
		\item Alice and Bob perform a verification protocol to check whether $\mathbf{A} = \hat{\mathbf{A}}$, through the use of almost-universal hashing (see Sec.~\ref{subsubsec:inforecon}). If the verification fails, they abort the protocol.
	\end{enumerate}
	\item Parameter estimation and (CHSH) acceptance test:
	\begin{enumerate}
		\item Bob announces the value of $b_i$ for every round $i$ that is a test round for both Alice and Bob, i.e. $t_{i}=s_{i}=0$. Alice uses these values to compute the value
		\begin{equation}
		\beta_\mathrm{freq}
		=\frac{\sum_{i:s_{i}=t_{i}=0}\omega(
			{a}_{i},b_{i},x_{i},y_{i})}{n}.
		\end{equation}
		\item If $\beta_\mathrm{freq}<\gamma_A\gamma_B\omega_{\mathrm{exp}}-\delta$, Alice and Bob abort the protocol.
	\end{enumerate}
	\item Privacy amplification:
	\begin{enumerate}
		\item Alice and Bob use a randomness extractor to generate final keys $\mathbf{K}_{A}$ and $\mathbf{K}_{B}$ of length $\ell$ from $\mathbf{A}$ and $\hat{\mathbf{A}}$.
	\end{enumerate}
\end{enumerate}

\subsubsection{QKD security}

For QKD protocols, security is usually defined in the following way: 
A QKD protocol is $(\epsilon_{\mathrm{com}},\epsilon_{\mathrm{snd}},\ell)$ secure if the protocol outputs a key of length $\ell$ that is $\epsilon_{\mathrm{com}}$ complete and $\epsilon_{\mathrm{snd}}$ sound.
Here, completeness describes that when the protocol is honestly implemented, the abort probability is less than $\epsilon_{\mathrm{com}}$. 

Soundness must hold against all attacks an adversary can perform, 
and states that the state produced at the end of the protocol is $\epsilon_{\mathrm{snd}}$-close (in trace distance) to a more ``ideal'' state; see e.g.~\cite{PR22,di_security_review2} for more details. It is often convenient in security proofs to decompose the soundness parameter into a correctness parameter $\epsilon_{\mathrm{cor}}$ and a secrecy parameter $\epsilon_{\mathrm{sec}}$ satisfying $\epsilon_{\mathrm{cor}}+\epsilon_{\mathrm{sec}}\leqslant\epsilon_{\mathrm{snd}}$, as described in~\cite{PR22}
--- qualitatively, correctness is an upper bound on the probability that the final keys differ and the protocol accepts, while secrecy quantifies how close the Alice-Eve state is to a more ``ideal'' state.

Different approaches and techniques exist to prove the information-theoretic security of the CHSH-based protocol~\cite{arnon2018practical,VaziraniVidick2014Fully,MillerShi2016Robust,ZhangEtAl2023Quantum}.
All these analyses prove that, 
in realistic scenarios, 
the experimental data (namely, $\mathbf{X}$, $\mathbf{A}$, $\mathbf{T}$, $\mathbf{Y}$, $\mathbf{B}$, and $\mathbf{S}$) yield a final key with length up to some value $\ell$.
On the other hand, a better approach can produce a larger $\ell$ given fixed $\epsilon_{\mathrm{com}}$ and $\epsilon_{\mathrm{snd}}$.
Below we outline two types of security analysis we used in the experiment.

\subsubsection{The original EAT approach}

The Entropy Accumulation Theorem (EAT)~\cite{DFR20} is a powerful result in quantum information theory that formalizes how entropy builds up over many sequential rounds of a process, even when the rounds are not independent and identically distributed.
In the context of DI-QKD, EAT has become a central tool because it enables rigorous finite-key security proofs that often predict significantly higher key rates than earlier approaches, bringing practical DI-QKD within reach of current or near-term experimental capabilities.
In essence, EAT states that if a process can be decomposed into a sequence of steps, each producing some amount of entropy conditioned on prior steps and an adversary's quantum side information, then the total entropy is approximately the sum of the per-round entropies, up to a finite-size correction.
In this work, we build on established EAT-based results~\cite{arnon2018practical,TSB+22} and adapt them to the particulars of our protocol, ensuring that the entropy analysis accurately reflects the sifting steps used in our implementation.

\paragraph{The completeness.} The completeness error $\epsilon_{\mathrm{com}}$ for the protocol satisfies
\begin{equation}
\epsilon_{\mathrm{com}}
\leqslant
\epsilon_{\mathrm{EC}}^{\mathrm{com}}
+\epsilon_{\mathrm{EA}}^{\mathrm{com}},
\end{equation}
where $\epsilon_{\mathrm{EC}}^{\mathrm{com}}$ is the probability that EC protocol yields $\hat{\mathbf{A}}\neq\mathbf{A}$ in the honest case, 
and $\epsilon_{\mathrm{EA}}^{\mathrm{com}}$ is the probability of failing the acceptance test in the honest case (due to statistical fluctuations).
Here, using Chernoff-Hoeffding inequality, we have
\begin{align}
\epsilon_{\mathrm{EA}}^{\mathrm{com}} & \leqslant e^{-nD\left[c\|\gamma_{A}\gamma_{B}\omega_{\mathrm{exp}}\right]},
\end{align}
where $D[p\|q]=p\log_{2}(p/q)+(1-p)\log_{2}[(1-p)/(1-q)]$.

\paragraph{The soundness.} The soundness error $\epsilon_{\mathrm{snd}}$ for the protocol satisfies
\begin{equation}
\epsilon_{\mathrm{snd}}\leqslant\epsilon_{\mathrm{EC}}+\epsilon_{\mathrm{PA}}+\epsilon_{s},
\end{equation}
arising from a correctness parameter of $\epsilon_{\mathrm{cor}} = \epsilon_{\mathrm{EC}}$ and a secrecy parameter of $\epsilon_{\mathrm{sec}} = \epsilon_{\mathrm{PA}}+ \epsilon_{s}$ (which we elaborate further on below), with the latter coming from the
privacy amplification step in the protocol.

Essentially, privacy amplification allows us to bound the trace distance between the real Alice-Eve state at the end of the protocol and a more ``ideal'' state, in terms of the difference between the final key length $\ell$ and the smooth min-entropy $H_{\min}^{\epsilon_{s}}\left(\mathbf{A}|\mathbf{O}E\right)$ of the state conditioned on accepting, where $\mathbf{O}$ stands for all the classical strings publicly communicated in the protocol (we discuss this in more detail later). 

\paragraph{The key length.} Given the above parameters, an achievable secret key length $\ell$ is
\begin{align}
\ell & = n\eta_{\mathrm{opt}}\left(\epsilon_{s}^{\prime},\epsilon_{\mathrm{EA}}+\epsilon_{\mathrm{EC}}\right)-\mathrm{leak}_{\mathrm{EC}}
-\mathrm{leak}_{\mathrm{EV}}
-2\vartheta_{\epsilon_{s}-\epsilon_{s}^{\prime}-2\epsilon_{s}^{\prime\prime}}\nonumber \\
& \quad-\gamma_{A}\gamma_{B}n-\sqrt{n}\log5\sqrt{1-2\log\left(\epsilon_{s}^{\prime\prime}\left(\epsilon_{\mathrm{EA}}+\epsilon_{\mathrm{EC}}\right)\right)}-2\log\frac{1}{\epsilon_{\mathrm{PA}}}.\label{eq:kr-AFRV}
\end{align}
Here, $\mathrm{leak}_{\mathrm{EC}}$ and $\mathrm{leak}_{\mathrm{EV}}$ correspond to the information leakage
	caused by the various steps during information reconciliation (which we elaborate on in Sec.~\ref{subsubsec:inforecon} later), $\epsilon_{s}^{\prime}$ and $\epsilon_{s}^{\prime\prime}$ are arbitrary parameters satisfying $\epsilon_{s}-\epsilon_{s}^{\prime}-2\epsilon_{s}^{\prime\prime}>0$ that are to be optimized over, $\vartheta_{\epsilon}$ is a function satisfying $\vartheta_{\epsilon}\leqslant1-2\log\epsilon$,
and 
\begin{gather}
\eta_{\mathrm{opt}}\left(\epsilon,\epsilon_{e}\right)=\max_{\frac{3}{4}<p_{t}<\frac{2+\sqrt{2}}{4}}\eta\left(\omega_{\mathrm{exp}}-\delta/\gamma_{\mathrm{eff}},p_{t},\epsilon,\epsilon_{e}\right),\\
\eta\left(\omega,\omega_{t},\epsilon,\epsilon_{e}\right)=\gamma_{\mathrm{eff}}f\left(\omega,\omega_{t}\right)-\frac{2}{\sqrt{n}}\left(\log9+\left\lceil \gamma_{\mathrm{eff}}\left.\frac{d}{dp}g\left(p\right)\right|_{p_{t}}\right\rceil \right)\sqrt{1-2\log\left(\epsilon\epsilon_{e}\right)},\\
f\left(\omega,\omega_{t}\right)=\begin{cases}
g\left(\omega\right), & p\leqslant p_{t},\\
\left.\frac{d}{dp}g\left(\omega\right)\right|_{\omega_{t}}\left(\omega-\omega_{t}\right)+g\left(\omega_{t}\right), & p>p_{t},
\end{cases}\label{eq:fmin-AFRV}\\
g\left(\omega\right)=\begin{cases}
1-h\left(\frac{1}{2}+\frac{1}{2}\sqrt{16\omega\left(\omega-1\right)+3}\right), & \omega\in\left(\frac{3}{4},\frac{2+\sqrt{2}}{4}\right),\\
\text{undefined,} & \text{otherwise,}
\end{cases}
\end{gather}
with $\gamma_{\mathrm{eff}}=1-\gamma_{A}/2-\gamma_{B}+3\gamma_{A}\gamma_{B}/2$
being the ratio of outcomes that were not deterministically set as $(0,0)$.

We briefly explain the modifications from previous literature.
The main difference between the current protocol and the ones proposed
in~\cite{arnon2018practical,TSB+22} is that the current one has the sifting procedure, i.e.,
some outcomes are sifted by setting them to $0$. (There is also a simplification in the parameter estimation step that we discuss later.) The sifting procedure affects
the so-called min-tradeoff function used for bounding the smooth min-entropy that is used to quantify the length of final keys. The min-tradeoff
function $f_{\min}$ should have certain properties and satisfy 
\begin{equation}
f_{\min}\left(p\right)\leqslant\inf_{i,\rho_{i}}H\left(A_{i}B'_{i}|X_{i}Y_{i}S_{i}T_{i}R^{\prime}\right)_{\rho_{i}},
\end{equation}
where $H$ is the von Neumann entropy,
$B'_{i}$ is a register that is equal to $B_{i}$ in test rounds and set to a fixed value otherwise,
$p$ is the probability distribution
over the testing result (i.e., $\omega(a_{i},b_{i},x_{i},y_{i})$
for testing rounds and $0$ for other rounds), $i$ labels the round,
$R^{\prime}$ is an arbitrary quantum system correlated with Alice's
and Bob's systems, and $\rho_{i}$ is an arbitrary state that can be
generated in a single round of the protocol while being ``compatible'' with the winning probability $p$ (see~\cite{arnon2018practical,TSB+22} for details). For the case without sifting, the min-tradeoff function
has been proven as Eq.~(\ref{eq:fmin-AFRV})~\cite{arnon2018practical}, and the essential
technical derivation is that
\begin{align}
H\left(A_{i}B'_{i}|X_{i}Y_{i}S_{i}T_{i}R^{\prime}\right)_{\rho_{i}} & \geqslant\sum_{x_{i}=0,1}P\left(x_{i}\right)H\left(A_{i}|X_{i}=x_{i},S_{i}T_{i}R^{\prime}\right)\\
& \geqslant f\left(\omega,\omega_{t}\right)
\end{align}
for any $\omega_{t}$. With sifting procedure, we modify it as
\begin{align}
H\left(A_{i}B'_{i}|X_{i}Y_{i}S_{i}T_{i}R^{\prime}\right)_{\rho_{i}} & =\gamma_{\mathrm{eff}}\sum_{i:\text{ not sifted}}H\left(A_{i}B'_{i}|X_{i}Y_{i}S_{i}T_{i}R^{\prime}\right)_{\rho_{i}}\nonumber \\
& \qquad+\left(1-\gamma_{\mathrm{eff}}\right)\sum_{i:\text{ sifted}}H\left(A_{i}B'_{i}|X_{i}Y_{i}S_{i}T_{i}R^{\prime}\right)_{\rho_{i}}\\
& \geqslant\gamma_{\mathrm{eff}}f\left(\omega,\omega_{t}\right),
\end{align}
i.e., $\gamma_{\mathrm{eff}}f\left(\omega,\omega_{t}\right)$ is a
min-tradeoff function for the protocol with sifting. 

Combining with the EAT, one can show that for any state that passes the acceptance test with probability larger than $\epsilon_{\mathrm{EA}}+\epsilon_{\mathrm{EC}}$, the state conditioned on accepting in that step satisfies
\begin{align}
H_{\min}^{\epsilon_{s}^{\prime}}\left(\mathbf{AB}'|\mathbf{XYST}E\right) & \geqslant n\eta_{\mathrm{opt}}\left(\epsilon_{s}^{\prime},\epsilon_{\mathrm{EA}}+\epsilon_{\mathrm{EC}}\right).
\end{align}
(We can restrict our attention to such states without loss of generality, because any state accepted in that step with lower probability than this is trivially $(\epsilon_{\mathrm{EA}}+\epsilon_{\mathrm{EC}})$-secret; see e.g.~\cite{arnon2018practical,TSB+22}.)
However, this is not yet enough to apply a privacy amplification theorem, since in the protocol there are other publicly announced strings such as $\bf{B}'$ (i.e.~Bob's test-round data), the error-correction string $\bf{L}$, the verification hash, and so on. To handle this, we apply chain rules~\cite{vitanov2013} for smooth entropies, which yield bounds such as
\begin{align}
H_{\min}^{\epsilon_{s}}\left(\mathbf{A}|\mathbf{B}'\mathbf{XYST}E\right)  
\geqslant
H_{\min}^{\epsilon_{s}^{\prime}}\left(\mathbf{AB}'|\mathbf{XYST}E\right) - H_{\max}^{\epsilon_{s}^{\prime\prime}}\left(\mathbf{B}'|\mathbf{XYST}E\right)  
-2\vartheta_{\epsilon_{s}-\epsilon_{s}^{\prime}-2\epsilon_{s}^{\prime\prime}},
\end{align}
and so on. Continuing on in a similar fashion to the analysis in~\cite{TSB+22} and applying a privacy amplification theorem,
we obtain the key-rate formula in
Eq.~(\ref{eq:kr-AFRV}). However, here we end up with a better soundness parameter of $\epsilon_{\mathrm{EC}}+\epsilon_{\mathrm{PA}}+\epsilon_{s}$ compared to the value $2\epsilon_{\mathrm{EC}}+\epsilon_{\mathrm{PA}}+\epsilon_{s}$ obtained in that work, because in our protocol we have Bob announce his test-round outcomes directly for parameter estimation; it was observed in~\cite{arq2024generalized} that this simplifies the security proof without notably affecting EAT-based key-rates.

\subsubsection{The R\'{e}nyi EAT approach}
We follow the security analysis provided in~\cite{hahn2025analytic} and extend it to incorporate the sifting step. We highlight that here we slightly modify the 
accept condition for the protocol, in order to obtain improvements in some bounds.

Let $c_{i} $ denote the following test information of the $i$'th round:
\begin{align}
c_{i} \coloneqq \begin{cases}
\omega(a_{i},b_{i},x_{i},y_{i}), & \text{if } s_i=t_i=0 \\
\perp, & \text{else} \; .
\end{cases}
\end{align}
This corresponds to $c_i$ outputting $\omega$ during test-rounds, and $\perp$ during rounds that are sifted out or during key-generating rounds. Let us now specify the set of observed behaviors under which the protocol accepts. For the honest behavior we can write
\begin{align}
\begin{aligned}
&q_{hon}(0) = \gamma_{A}\gamma_{B} \left(1-\omega_{\mathrm{exp}}\right)\; , \qquad q_{hon}(1) = \gamma_{A}\gamma_{B} \omega_{\mathrm{exp}}\; , \qquad
&q_{hon}(\perp) = 1-\gamma_{A}\gamma_{B} \; .
\end{aligned}
\end{align}
We now choose the accept condition to be that the observed frequency distribution on the $\mathbf{C}$ registers (which we denote by $\freq_{\mathbf{C}}$) lies in a set $S_{acc}$, defined as the set of all distributions $q$ such that
\begin{align}
\forall c \in \{0,1,\perp\}, \quad 
q_\mathrm{hon}(c) - \dlow_{c} \leq q(c) \leq q_\mathrm{hon}(c) + \dupp_{c}
\; ,
\end{align}
where $\dlow_{c},\dupp_{c}$ are chosen such that the honest behavior satisfies
\begin{align}
\forall c \in \{0,1,\perp\}, &\quad \Pr[\freq_{\mathbf{C}}(c) < q_\mathrm{hon}(c) - \dlow_{c}] \leq \frac{\ecomAT}{6} 
\quad\\\text{and}&\quad\Pr[\freq_{\mathbf{C}}(c) > q_\mathrm{hon}(c) + \dupp_{c}] \leq \frac{\ecomAT}{6}  \;
,
\end{align}
where $\ecomAT$ is the (acceptance test) completeness parameter. As discussed in~\cite{hahn2025analytic}, this ensures an honest behavior aborts the acceptance test at most $\ecomAT$ of the time. We construct $\dlow_{c},\dupp_{c}$ numerically, in the same vein as~\cite{hahn2025analytic}.

\newcommand{\eventAT}{\Omega_{EA}}
\paragraph{R\'{e}nyi EAT and Single-Round Analysis}
After all the measurement rounds of the protocol have been concluded, the global state is of the form
\begin{align}
\rho_{\mathbf{A}\mathbf{B}\mathbf{C}\mathbf{X}\mathbf{Y}\mathbf{S}\mathbf{T}\mathbf{E}} = \Pr[\eventAT]\rho_{|\eventAT} + \left(1-\Pr[\eventAT] \right) \rho_{\perp} \; ,
\end{align}
where $\rho_{|\eventAT}$ corresponds to the global state 
conditioned on the acceptance test passing.
Using~\cite[Lemma $5.1$]{arq2024generalized}, the R\'{e}nyi EAT provides bounds on the total accumulated R\'{e}nyi entropy of the form
\begin{align}
\widetilde{H}_{\alpha}^{\uparrow}\left(\mathbf{A}\mathbf{C}|
\mathbf{X}\mathbf{Y}\mathbf{S}\mathbf{T}\mathbf{E}\right)_{\rho_{|\eventAT}} \geq n h_{\alpha} 
- \frac{\alpha}{\alpha-1} \log\frac{1}{\Pr[\eventAT]} \; ,
\end{align}
where $h_{\alpha} $ is a single-round quantity we describe below. 
Critically, for any single round, Alice's and Bob's measurements can be modeled by POVM operators $\mathcal{M}=\{M_{a|x},N_{b|y} \}$, where $M_{a|x}$ ($N_{b|y}$) corresponds to the POVM element associated with the event that Alice (Bob) applies measurement $X=x$ ($Y=y$) and records the output $A=a$ ($B=b$). 

For this approach, the single-round quantity is given by
\begin{align} \label{eq: halphaREAT}
h_{\alpha} \coloneqq& \inf_{\mathcal{M}(\sigma)} \inf_{\vect{q} \in S_\mathrm{acc}}   \frac{1}{\alpha-1}D\left(
\vect{q} \middle\Vert \vect{p}_{\mathcal{M}(\sigma)}
\right) + q(\perp)\widetilde{H}_{\alpha}^{\downarrow}\left( A|C=\perp,XYSTE\right)_{\mathcal{M}(\sigma)} \; ,
\end{align}
where $\sigma$ denotes any initial state shared by Alice, Bob, and Eve, $\mathcal{M}$ denotes any single-round channel that Alice and Bob can apply on $\sigma$ to output the registers $ABCXYST$, and $\vect{p}_{M(\sigma)}$ denotes the test data generated by $\mathcal{M}$ acting on $\sigma$.
Using~\cite[Proposition $5.1$]{tomamichel2013framework}, which allows us to first separate the R\'{e}nyi entropy into the sifting and key-generation contributions and subsequently bound the corresponding sifting round contribution by $0$, we find that
\begin{multline}
\widetilde{H}_{\alpha}^{\downarrow}\left( A|C=\perp,XYSTE\right)_{\mathcal{M}(\sigma)} \geq \frac{1}{1-\alpha} \log \left( \frac{1- \gamma_{B} - \frac{1}{2}\gamma_{A}\left(1-\gamma_{B}\right) }{1-\gamma_{A}\gamma_{B}}2^{\left(1-\alpha\right)\widetilde{H}_{\alpha}^{\downarrow}\left( A|X=0, E\right)_{\mathcal{M}(\sigma)}} \right.\\
+ \left. \frac{\left(1-\gamma_{A}\right)\gamma_{B} + \frac{1}{2}\gamma_{A}\left(1-\gamma_{B}\right) }{1-\gamma_{A}\gamma_{B}}\right)\; ,
\end{multline}
where we additionally use the fact that Alice always applies the measurement $X=0$ during key-generating measurements. 

The explicit calculation for the single-round analysis is handled as follows. By construction, the single-round channel $\mathcal{M}$ acting on an arbitrary state $\sigma$ will generate a test-data distribution of the form
\begin{align} \label{Eq: SingleRoundPExpression1}
\vect{p}_{\mathcal{M}(\sigma)} \coloneqq \begin{pmatrix}
p_{\mathcal{M}(\sigma)} (0) \\
p_{\mathcal{M}(\sigma)} (1) \\
p_{\mathcal{M}(\sigma)} (\perp) 
\end{pmatrix}= \begin{pmatrix}
\gamma_{A}\gamma_{B} \left(1-\omega_{\sigma}\right) \\
\gamma_{A}\gamma_{B} \omega_{\sigma} \\
1-\gamma_{A}\gamma_{B} 
\end{pmatrix} \;, 
\end{align}
where $\omega_{\sigma}=\frac{1+S_{\sigma}/4}{2}$ is the CHSH winning probability achieved by $\mathcal{M}$ acting on $\sigma$ and $S_{\sigma}$ is the corresponding CHSH score. By~\cite{hahn2025analytic}, $\widetilde{H}_{\alpha}^{\downarrow}\left( A|X=0,E\right)_{\mathcal{M}(\sigma)}$ can be tightly bounded and expressed in terms of $S_{\sigma}$, i.e. the strongest single-round attack at a given value of $S_{\sigma}$ satisfies
\begin{align} \label{Eq: SingleRoundEntExpression1}
2^{\left(1-\alpha\right)\widetilde{H}_{\alpha}^{\downarrow}\left( A|X=0, E\right)_{\mathcal{M}(\sigma)}} = 2^{1-\alpha} \cdot \left[ \left(\frac{1-\sqrt{\frac{S_{\sigma}^{2}}{4}-1}}{2} \right)^{\frac{1}{\alpha}}+ \left(\frac{1+\sqrt{\frac{S_{\sigma}^{2}}{4}-1}}{2} \right)^{\frac{1}{\alpha}}\right]^\alpha \; .
\end{align}
Inputting the expressions given by Eqs.~\eqref{Eq: SingleRoundPExpression1}--\eqref{Eq: SingleRoundEntExpression1} reduces the single-round expression $h_{\alpha}$  to a $3$-variable optimization problem, which is convex in $\vect{q}$ and $S_{\sigma}$ individually and can be solved heuristically.

\paragraph{Key Length}
Following the analysis of~\cite{hahn2025analytic}, a feasible secure key length (with correctness $\epsilon_{\mathrm{cor}} = \epsilon_{\mathrm{EC}}$ and secrecy $\epsilon_{\mathrm{sec}}$) 
is given by 
\begin{align}
\ell = n h_{\alpha} - n\left(\gamma_{A}\gamma_{B} + \dlow_{\perp} \right) - \mathrm{leak}_\mathrm{EC} - \mathrm{leak}_\mathrm{EV} - \frac{\alpha}{\alpha-1}\log\frac{1}{\epsilon_{\mathrm{sec}}} + 2 \; ,
\end{align}

where again $\mathrm{leak}_\mathrm{EC}$ and $ \mathrm{leak}_\mathrm{EV}$ are as described in Sec.~\ref{subsubsec:inforecon} below.

\subsubsection{The information leakage for the information reconciliation.} \label{subsubsec:inforecon}
\begin{figure}[htbp]
	\centering
	\includegraphics[width=.9\linewidth]{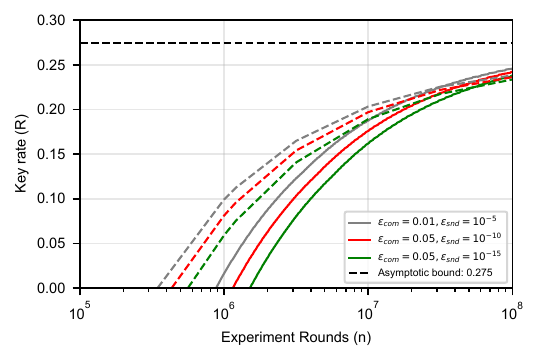}
	\caption{
		Simulation results for the original and R\'{e}nyi EAT. The solid lines correspond to the original EAT approach, while the dashed lines correspond to the R\'enyi EAT approach.
	}
	\label{fig:simulation}
\end{figure}
In both approaches, we use the same bound for the information leakage during the information reconciliation.
Specifically, we follow~\cite{TSB+22} and use a bound from~\cite{Renes_2012} (combined with~\cite[Corollary~4.10]{DFR20}):
\begin{equation}\label{eq:leakEC}
\mathrm{leak}_{\mathrm{EC}}\leqslant n\eta^{\infty}+2\log5\sqrt{n\log\frac{2}{\left(\epsilon_{\mathrm{EC}}^{\mathrm{com}}-
		\tilde{\epsilon}
		\right)^{2}}}+2\log\frac{1}{
	\tilde{\epsilon}
}+4,
\end{equation}
where $\tilde{\epsilon} \in [0,\epsilon_{\mathrm{EC}}^{\mathrm{com}})$ is an arbitrary value that can be optimized over, and $\eta^{\infty}$ is the rate of an optimal EC code in the
asymptotic limit, i.e.,
\begin{equation}
\eta^{\infty}=\left(1-\frac{\gamma_{A}}{2}\right)\left(1-\gamma_{B}\right)h\left(q\right)+\gamma_{A}\gamma_{B}h\left(\omega\right).
\end{equation}
Here, $q$ is the honest QBER for $(x,y)=(0,2)$ and $\omega$ is
the honest average winning probability for $(x,y)\in\{0,1\}^2$.

Meanwhile, for the error verification step, we follow~\cite{nadlinger2022experimental} and use an $\epsilon_{\mathrm{EC}}$-almost-universal hash of length 64 bits with $\epsilon_{\mathrm{EC}}=2^{-61}$ (specifically, the VHASH hash family, which is applicable to raw data of size up to $2^{64}$ bits). As discussed in that work, this yields
\begin{align}
\mathrm{leak}_{\mathrm{EV}}=64, \qquad
\epsilon_{\mathrm{cor}} = \epsilon_{\mathrm{EC}} = 2^{-61}.
\end{align}

\subsubsection{The simulation}

\begin{figure}[htbp]
	\centering
	\includegraphics[width=.9\linewidth]{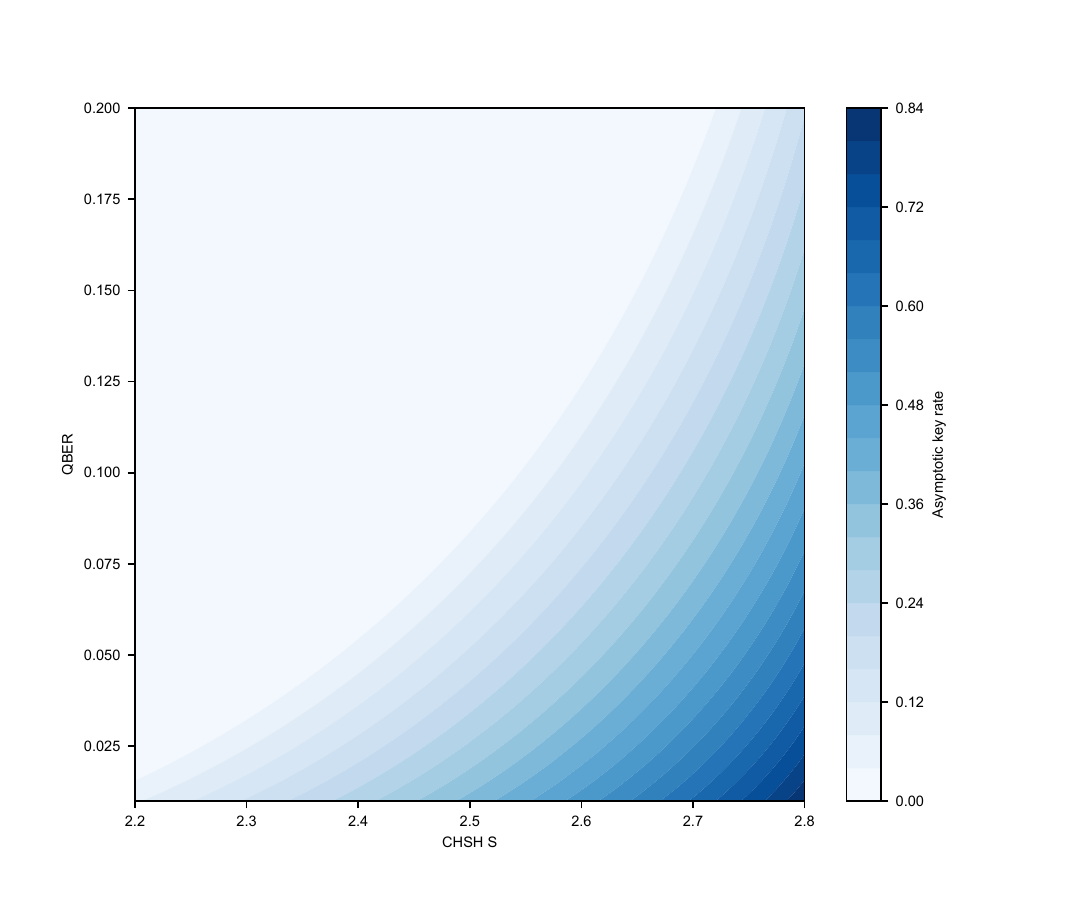}
	\caption{
		Contour plot of the Asymptotic key rate versus the CHSH parameter $S$ and the quantum bit error rate (QBER). The fraction of Bell tests is set to $\gamma_{A} =\gamma_{B} = 10^{-3}$. 
	}
	\label{fig:asymptotic}
\end{figure}

In this section, we provide a simulation of the key rates achievable by the two approaches discussed above.
We are interested in the parameters achieved in experiment, namely,
\begin{equation}
S = 2.612,\qquad Q = 0.0285,
\end{equation}
for $\gamma_A=0.26$ and $\gamma_B=0.13$. 
We set the security parameters as $\epsilon_{\mathrm{com}}=0.01, 0.05$ and $\epsilon_{\mathrm{snd}}=10^{-5}, 10^{-10}, 10^{-15}$.
The numerical results are shown in Fig.~\ref{fig:simulation}.  We highlight that the R\'{e}nyi EAT significantly outperforms the original EAT approach in the parameter regime that is of practical interest for the DI-QKD experiment. 
Moreover, the single-round quantity $h_{\alpha}$ from Eq.~\eqref{eq: halphaREAT} can further be improved upon in a DI setting using known results from~\cite{arq2024generalized,hahn2024boundspetzrenyidivergencesapplications}. One drawback of Eq.~\eqref{eq: halphaREAT} is that it discards entropy contributions from the test rounds. This can be mitigated, however, by either using the tighter single-round DI bounds found in~\cite{arq2024generalized,hahn2024boundspetzrenyidivergencesapplications} or by choosing smaller test-round fractions $\gamma_{A}, \gamma_{B}$ for large $n$. The fact that $\gamma_{A}, \gamma_{B}$ are fixed and independent of the number of rounds is the primary reason that the R\'{e}nyi EAT no longer outperforms the EAT for large $n$ in Fig.~\ref{fig:simulation}.

For the asymptotic key rate over long distances, we consider the DI-QKD protocol studied in ~\cite{arnon2018practical, nadlinger2022experimental}  which does not contain the sifting procedure. By setting the ratio of Bell-test rounds as $\gamma=0.001$ (and a ratio $1-\gamma$ of the experimental rounds are all used to generate keys), we obtain the key rate against QBER and CHSH value, as shown in Fig.~\ref{fig:asymptotic}.

\subsection{CHSH test}
{
	In this work, the measurement settings were chosen to align with the assumption of a maximally entangled two-qubit state. While optimizing the measurement bases could potentially improve the reported CHSH value, we opted for fixed measurement settings to maintain consistency across various experimental setups, particularly at different distances.
	
	In the 11\,km experiment, the dominant error sources were double excitation and residual photon recoil effects. These errors led to a maximum correlator along the Z direction, with measured values of $\langle \hat{Z} \hat{Z} \rangle = 0.943$ and $\langle \hat{X} \hat{X} \rangle = 0.924$. Given this anisotropy in the state, we deliberately chose to perform all Bell tests in the real-valued X-O-Z plane of the Bloch sphere rather than the X-O-Y plane.
	
	Theoretically, the optimal measurement bases for our actual state would be Alice's bases: $\{\hat{X},\hat{Z}\}$, and Bob's bases: $\{0.700\hat{X}+0.714\hat{Z},-0.700\hat{X}+0.714\hat{Z}\}$, which are close to the near-symmetric bases we used in practice. These error sources and the resulting anisotropy persist at longer distances. For consistency, we maintained identical measurement bases across all distances.
	
	To rigorously assess the statistical significance of our results, we conduct a p-value test under the null hypothesis to analyze the CHSH test data at each distance, as shown in Table~\ref{table:pvalue}. 
	
	The p-values are calculated using the binomial test to assess whether the observed winning probability in the CHSH game exceeds the classical limit of 0.75. The null hypothesis assumes that the winning probability per trial is 0.75, consistent with the optimal classical strategy. The game is modeled as a sequence of N Bernoulli trials, where the number of observed wins, $k$, follows a binomial distribution with parameters $N$ (number of trials) and $p$=0.75 (the probability of winning each trial). The $p$-value then describes the probability of observing $k$ wins or more extreme results under the assumption that the true winning probability is 0.75 according to local hidden variable theories.The results demonstrate that the probability of the observed data being consistent with classical correlations under the null hypothesis is exceedingly small, thereby providing evidence for the nonlocal nature of the measured correlations.

	\begin{table}[b]
	\centering
	\caption{\textbf{P-value for each fiber length and corresponding number of trials.}}
	\label{table:pvalue}
	\begin{tabular}{|c|c|c|}
		\hline
		\textbf{Fiber Length (km)} & \textbf{Number of Trials (N)} & \textbf{p-value} \\
		\hline
		11 & 39645 & 5.76$\times10^{-316}$ \\
		20 & 4351  & 1.92$\times10^{-28}$  \\
		50 & 1193  & 5.94$\times10^{-08}$  \\
		70 & 1166  & 5.50$\times10^{-07}$  \\
		100 & 920  & 4.72$\times10^{-06}$  \\
		\hline
	\end{tabular}
\end{table}
}


\begin{thebibliography}{10}
\providecommand{\url}[1]{\texttt{#1}}
\expandafter\ifx\csname urlstyle\endcsname\relax
  \providecommand{\doi}[1]{doi:\discretionary{}{}{}#1}\else
  \providecommand{\doi}{doi:\discretionary{}{}{}\begingroup
  \urlstyle{rm}\Url}\fi

\bibitem{wehner2018quantum}
S.~Wehner, D.~Elkouss, R.~Hanson, Quantum internet: A vision for the road
  ahead. \emph{Science} \textbf{362}, eaam9288 (2018).

\bibitem{xu_secure_2020}
F.~Xu, X.~Ma, Q.~Zhang, H.-K. Lo, J.-W. Pan, Secure quantum key distribution
  with realistic devices. \emph{Reviews of Modern Physics} \textbf{92}, 025002
  (2020).

\bibitem{pironio2009device}
S.~Pironio, \emph{et~al.}, Device-independent quantum key distribution secure
  against collective attacks. \emph{New Journal of Physics} \textbf{11}, 045021
  (2009).

\bibitem{arnon2018practical}
R.~Arnon-Friedman, F.~Dupuis, O.~Fawzi, R.~Renner, T.~Vidick, Practical
  device-independent quantum cryptography via entropy accumulation.
  \emph{Nature communications} \textbf{9}, 459 (2018).

\bibitem{TSB+22}
E.~Y.-Z. Tan, \emph{et~al.}, {Improved {DIQKD} protocols with finite-size
  analysis}. \emph{{Quantum}} \textbf{6}, 880 (2022).

\bibitem{Ekert1991Quantum}
A.~K. Ekert, Quantum cryptography based on {Bell}'s theorem. \emph{Physical
  Review Letters} \textbf{67}, 661--663 (1991), publisher: American Physical
  Society.

\bibitem{nadlinger2022experimental}
D.~P. Nadlinger, \emph{et~al.}, Experimental quantum key distribution certified
  by Bell's theorem. \emph{Nature} \textbf{607}, 682--686 (2022).

\bibitem{zhang2022device}
W.~Zhang, \emph{et~al.}, A device-independent quantum key distribution system
  for distant users. \emph{Nature} \textbf{607}, 687--691 (2022).

\bibitem{liu2022toward}
W.-Z. Liu, \emph{et~al.}, Toward a photonic demonstration of device-independent
  quantum key distribution. \emph{Physical Review Letters} \textbf{129}, 050502
  (2022).

\bibitem{PhysRevLett.105.070501}
N.~Gisin, S.~Pironio, N.~Sangouard, Proposal for Implementing
  Device-Independent Quantum Key Distribution Based on a Heralded Qubit
  Amplifier. \emph{Physical Review Letters} \textbf{105}, 070501 (2010).

\bibitem{yu2020entanglement}
Y.~Yu, \emph{et~al.}, Entanglement of two quantum memories via fibres over
  dozens of kilometres. \emph{Nature} \textbf{578}, 240--245 (2020).

\bibitem{van2022entangling}
T.~Van~Leent, \emph{et~al.}, Entangling single atoms over 33 km telecom fibre.
  \emph{Nature} \textbf{607}, 69--73 (2022).

\bibitem{liu2024creation}
J.-L. Liu, \emph{et~al.}, Creation of memory--memory entanglement in a
  metropolitan quantum network. \emph{Nature} \textbf{629}, 579--585 (2024).

\bibitem{knaut2024entanglement}
C.~M. Knaut, \emph{et~al.}, Entanglement of nanophotonic quantum memory nodes
  in a telecom network. \emph{Nature} \textbf{629}, 573--578 (2024).

\bibitem{stolk2024metropolitan}
A.~J. Stolk, \emph{et~al.}, Metropolitan-scale heralded entanglement of
  solid-state qubits. \emph{Science advances} \textbf{10} (2024).

\bibitem{zaske2012visible}
S.~Zaske, \emph{et~al.}, Visible-to-telecom quantum frequency conversion of
  light from a single quantum emitter. \emph{Physical Review Letters}
  \textbf{109}, 147404 (2012).

\bibitem{de2012quantum}
K.~De~Greve, \emph{et~al.}, Quantum-dot spin--photon entanglement via frequency
  downconversion to telecom wavelength. \emph{Nature} \textbf{491}, 421--425
  (2012).

\bibitem{cabrillo1999creation}
C.~Cabrillo, J.~I. Cirac, P.~Garcia-Fernandez, P.~Zoller, Creation of entangled
  states of distant atoms by interference. \emph{Physical Review A}
  \textbf{59}, 1025 (1999).

\bibitem{levine2022dispersive}
H.~Levine, \emph{et~al.}, Dispersive optical systems for scalable Raman driving
  of hyperfine qubits. \emph{Physical Review A} \textbf{105}, 032618 (2022).

\bibitem{SM}
Materials and methods are available as supplementary materials.

\bibitem{krutyanskiy2023entanglement}
V.~Krutyanskiy, \emph{et~al.}, Entanglement of trapped-ion qubits separated by
  230 meters. \emph{Physical Review Letters} \textbf{130}, 050803 (2023).

\bibitem{ruskuc2025multiplexed}
A.~Ruskuc, \emph{et~al.}, Multiplexed entanglement of multi-emitter quantum
  network nodes. \emph{Nature} \textbf{639}, 54--59 (2025).

\bibitem{humphreys2018deterministic}
P.~C. Humphreys, \emph{et~al.}, Deterministic delivery of remote entanglement
  on a quantum network. \emph{Nature} \textbf{558}, 268--273 (2018).

\bibitem{yang2025entangling}
C.-W. Yang, \emph{et~al.}, Entangling Two Rydberg Superatoms via Single-Photon
  Interference. \emph{Physical Review Letters} \textbf{135}, 110802 (2025).

\bibitem{slodivcka2013atom}
L.~Slodi{\v{c}}ka, \emph{et~al.}, Atom-atom entanglement by single-photon
  detection. \emph{Physical Review Letters} \textbf{110}, 083603 (2013).

\bibitem{dudin2012strongly}
Y.~Dudin, A.~Kuzmich, Strongly interacting Rydberg excitations of a cold atomic
  gas. \emph{Science} \textbf{336}, 887--889 (2012).

\bibitem{fedoseev2025coherent}
V.~Fedoseev, \emph{et~al.}, Coherent and Incoherent Light Scattering by
  Single-Atom Wave Packets. \emph{Physical Review Letters} \textbf{135}, 043601
  (2025).

\bibitem{zhang2025Tunable}
Y.-C. Zhang, \emph{et~al.}, Tunable Einstein-Bohr Recoiling-Slit
  Gedankenexperiment at the Quantum Limit. \emph{Physical Review Letters}
  \textbf{135}, 230202 (2025).

\bibitem{guhne2009}
O.~G{\"u}hne, G.~T{\'o}th, Entanglement detection. \emph{Physics Reports}
  \textbf{474}, 1--75 (2009).

\bibitem{arq2024generalized}
A.~Arqand, T.~A. Hahn, E.~Y.-Z. Tan, Generalized R\'enyi Entropy Accumulation
  Theorem and Generalized Quantum Probability Estimation. \emph{Physical Review
  X} \textbf{15}, 041013 (2025).

\bibitem{MayersYao1998Quantum}
D.~Mayers, A.~Yao, Quantum cryptography with imperfect apparatus, in
  \emph{Proceedings 39th {Annual} {Symposium} on {Foundations} of {Computer}
  {Science} ({Cat}. {No}.{98CB36280})} (1998), pp. 503--509, iSSN: 0272-5428.

\bibitem{vsupic2020self}
I.~{\v{S}}upi{\'c}, J.~Bowles, Self-testing of quantum systems: a review.
  \emph{Quantum} \textbf{4}, 337 (2020).

\bibitem{hensen2015loophole}
B.~Hensen, \emph{et~al.}, Loophole-free Bell inequality violation using
  electron spins separated by 1.3 kilometres. \emph{Nature} \textbf{526},
  682--686 (2015).

\bibitem{rosenfeld2017event}
W.~Rosenfeld, \emph{et~al.}, Event-ready Bell test using entangled atoms
  simultaneously closing detection and locality loopholes. \emph{Physical
  Review Letters} \textbf{119}, 010402 (2017).

\bibitem{giustina2015significant}
M.~Giustina, \emph{et~al.}, Significant-loophole-free test of Bell’s theorem
  with entangled photons. \emph{Physical Review Letters} \textbf{115}, 250401
  (2015).

\bibitem{shalm2015strong}
L.~K. Shalm, \emph{et~al.}, Strong loophole-free test of local realism.
  \emph{Physical Review Letters} \textbf{115}, 250402 (2015).

\bibitem{ikuta2011wide}
R.~Ikuta, \emph{et~al.}, Wide-band quantum interface for
  visible-to-telecommunication wavelength conversion. \emph{Nature
  communications} \textbf{2}, 537 (2011).

\bibitem{van2020long}
T.~van Leent, \emph{et~al.}, Long-distance distribution of atom-photon
  entanglement at telecom wavelength. \emph{Physical Review Letters}
  \textbf{124}, 010510 (2020).

\bibitem{petrovich2025broadband}
M.~Petrovich, \emph{et~al.}, Broadband optical fibre with an attenuation lower
  than 0.1 decibel per kilometre. \emph{Nature Photonics} \textbf{19},
  1203--1208 (2025).

\bibitem{hartung2024quantum}
L.~Hartung, M.~Seubert, S.~Welte, E.~Distante, G.~Rempe, A quantum-network
  register assembled with optical tweezers in an optical cavity. \emph{Science}
  \textbf{385}, 179--183 (2024).

\bibitem{li2025parallelized}
L.~Li, \emph{et~al.}, Parallelized telecom quantum networking with a
  ytterbium-171 atom array. \emph{Nature Physics} \textbf{21}, 1826--1833
  (2025).

\bibitem{Bennett96}
C.~H. Bennett, \emph{et~al.}, Purification of Noisy Entanglement and Faithful
  Teleportation via Noisy Channels. \emph{Physical Review Letters} \textbf{76},
  722--725 (1996).

\bibitem{pan_entanglement_2001}
J.-W. Pan, C.~Simon, C.~Brukner, A.~Zeilinger, Entanglement purification for
  quantum communication. \emph{Nature} \textbf{410}, 1067--1070 (2001).

\bibitem{pan_experimental_2003}
J.-W. Pan, S.~Gasparoni, R.~Ursin, G.~Weihs, A.~Zeilinger, Experimental
  entanglement purification of arbitrary unknown states. \emph{Nature}
  \textbf{423}, 417--836 (2003).

\bibitem{kalb2017entanglement}
N.~Kalb, \emph{et~al.}, Entanglement distillation between solid-state quantum
  network nodes. \emph{Science} \textbf{356}, 928--932 (2017).

\bibitem{evered2023high}
S.~J. Evered, \emph{et~al.}, High-fidelity parallel entangling gates on a
  neutral-atom quantum computer. \emph{Nature} \textbf{622}, 268--272 (2023).

\bibitem{azuma_quantum_2023}
K.~Azuma, \emph{et~al.}, Quantum repeaters: {From} quantum networks to the
  quantum internet. \emph{Reviews of Modern Physics} \textbf{95}, 045006
  (2023).

\bibitem{Zenodo}
B.-W. Lu, \emph{et~al.}, Data for "Device-independent quantum key distribution
  over 100 km with single atoms", version 1, Zenodo (2025),
  \doi{10.5281/zenodo.17958512}.

\bibitem{thompson2013coherence}
J.~D. Thompson, T.~Tiecke, A.~S. Zibrov, V.~Vuleti{\'c}, M.~D. Lukin, Coherence
  and Raman sideband cooling of a single atom in an optical tweezer.
  \emph{Physical Review Letters} \textbf{110}, 133001 (2013).

\bibitem{luo2025entangling}
X.-Y. Luo, \emph{et~al.}, Entangling quantum memories over 420 km in fiber
  (2025), arXiv:2504.05660.

\bibitem{CHSH}
J.~Clauser, M.~Horne, A.~Shimony, R.~Holt, {Proposed Experiment to Test Local
  Hidden-Variable Theories}. \emph{Physical Review Letters} \textbf{23},
  880--884 (1969).

\bibitem{PR22}
C.~Portmann, R.~Renner, {Security in quantum cryptography}. \emph{Reviews of
  Modern Physics} \textbf{94}, 025008 (2022).

\bibitem{di_security_review2}
I.~W. Primaatmaja, \emph{et~al.}, {Security of device-independent quantum key
  distribution protocols: a review}. \emph{Quantum} \textbf{7}, 932 (2023).

\bibitem{VaziraniVidick2014Fully}
U.~Vazirani, T.~Vidick, Fully {Device}-{Independent} {Quantum} {Key}
  {Distribution}. \emph{Physical Review Letters} \textbf{113}, 140501 (2014).

\bibitem{MillerShi2016Robust}
C.~A. Miller, Y.~Shi, Robust {Protocols} for {Securely} {Expanding}
  {Randomness} and {Distributing} {Keys} {Using} {Untrusted} {Quantum}
  {Devices}. \emph{Journal of the ACM} \textbf{63}, 33:1--33:63 (2016).

\bibitem{ZhangEtAl2023Quantum}
X.~Zhang, P.~Zeng, T.~Ye, H.-K. Lo, X.~Ma, Quantum {Complementarity} {Approach}
  to {Device}-{Independent} {Security}. \emph{Physical Review Letters}
  \textbf{131}, 140801 (2023).

\bibitem{DFR20}
F.~Dupuis, O.~Fawzi, R.~Renner, {Entropy Accumulation}. \emph{Communications in
  Mathematical Physics} \textbf{379}, 867--913 (2020).

\bibitem{vitanov2013}
A.~Vitanov, F.~Dupuis, M.~Tomamichel, R.~Renner, {Chain Rules for Smooth Min-
  and Max-Entropies}. \emph{IEEE Transactions on Information Theory}
  \textbf{59}, 2603--2612 (2013).

\bibitem{hahn2025analytic}
T.~A. Hahn, A.~Philip, E.~Y.-Z. Tan, P.~Brown, Analytic R\'enyi Entropy Bounds
  for Device-Independent Cryptography  (2025), arXiv:2507.07365.

\bibitem{tomamichel2013framework}
M.~Tomamichel, A Framework for Non-Asymptotic Quantum Information Theory
  (2013), arXiv:1203.2142.

\bibitem{Renes_2012}
J.~M. Renes, R.~Renner, One-Shot Classical Data Compression With Quantum Side
  Information and the Distillation of Common Randomness or Secret Keys.
  \emph{IEEE Transactions on Information Theory} \textbf{58}, 1985–1991
  (2012).

\bibitem{hahn2024boundspetzrenyidivergencesapplications}
T.~A. Hahn, E.~Y.~Z. Tan, P.~Brown, {Bounds on Petz-R\'enyi Divergences and
  their Applications for Device-Independent Cryptography}  (2024),
  arXiv:2408.12313.

\end{thebibliography}
\end{document}